\documentclass[10pt,journal,comsoc]{IEEEtran}

\ifCLASSOPTIONcompsoc
\usepackage[nocompress]{cite}
\else
\usepackage{cite}
\fi

\ifCLASSINFOpdf

\else

\fi

\usepackage{url}
\usepackage{cite}
\usepackage{amsmath,amssymb,amsfonts}
\usepackage[ruled,noend,linesnumbered]{algorithm2e} %

\usepackage{algpseudocode}
\usepackage{graphicx}
\usepackage{xcolor}%
\usepackage{graphicx}
\usepackage{textcomp}
\usepackage{enumitem}
\usepackage{url}
\usepackage{amsmath}
\usepackage{multirow}
\usepackage{threeparttable}
\usepackage{verbatim}
\usepackage{booktabs}
\usepackage{array}

\definecolor{red}{rgb}{0,0,0} %
\definecolor{blue}{rgb}{0,0,0} %
\definecolor{blue-v2}{rgb}{0,0,0} %

\definecolor{dispatchers1}{rgb}{0.412,0.392,0.482} %
\definecolor{orchestrator1}{rgb}{0.69,0.478,0.596} %
\definecolor{dispatchers2}{rgb}{0.380,0.227,0.545} %
\definecolor{orchestrator2}{rgb}{0.525,0.220,0.271} %

\newcommand{\vthird}[1]{{\color{blue-v2} #1}}

\begin{document}

\title{Collaborative Learning-Based Scheduling for \textit{Kubernetes}-Oriented Edge-Cloud Network}

\author{Shihao~Shen,~\IEEEmembership{Student Member,~IEEE,}
		Yiwen~Han,~\IEEEmembership{Student Member,~IEEE,}
		Xiaofei~Wang,~\IEEEmembership{Senior Member,~IEEE,}
        Shiqiang~Wang,~\IEEEmembership{Member,~IEEE,} 
        and~Victor~C.M.~Leung,~\IEEEmembership{Life~Fellow,~IEEE}%

\thanks{ \rm{ %
Manuscript received 19 April 2021; revised 22 July 2022 and 23 March 2023; accepted 2 April 2023.
Xiaofei Wang was supported by the National Science Foundation of China (Grant 62072332), the China NSFC (Youth) (Grant 62002260), the China Postdoctoral Science Foundation (Grant 2020M670654), and the Tianjin Xinchuang Haihe Lab (Grant 22HHXCJC00002).
Victor C.M. Leung was supported by the Guangdong Pearl River Talent Recruitment Program (Grant 2019ZT08X603), the Guangdong Pearl River Talent Plan (Grant  2019JC01X235), Shenzhen Science and Technology Innovation Commission (Grant R2020A045), and the Canadian Natural Sciences and Engineering Research Council (Grant RGPIN-2019-06348).
A preliminary version of this paper titled ``Tailored Learning-Based Scheduling for Kubernetes-Oriented Edge-Cloud System'' was presented in the IEEE International Conference on Computer Communications (INFOCOM), 2021~\cite{kais}.
This paper extends the previous work by enhancing the detailed description, refining the system design, adding the fine-grained experiments about dequeue strategies and encoding methods.
} \textit{(Corresponding author: Xiaofei Wang})
    
\rm{Shihao Shen, Yiwen Han and Xiaofei Wang are with the College of Intelligence and Computing, Tianjin University, Tianjin 300350, China (e-mail: \{shenshihao, hanyiwen, xiaofeiwang\}@tju.edu.cn).
  
Shiqiang Wang is with IBM T. J. Watson Research Center, Yorktown Heights, NY 10598, USA (e-mail: shiqiang.wang@ieee.org).
  
Victor C.M. Leung is with the College of Computer Science and Software Engineering, Shenzhen University, Shenzhen 518052, China, and also with the Department of Electrical and Computer Engineering, The University of British Columbia, Vancouver, Canada V6T 1Z4 (e-mail: vleung@ieee.org).}}
}

\markboth{IEEE/ACM TRANSACTIONS ON NETWORKING,~Vol.~XX, No.~X, April~2023}
{Shen \MakeLowercase{\textit{et al.}}: Collaborative Learning-Based Scheduling for \textit{Kubernetes}-Oriented Edge-Cloud Network}

\IEEEtitleabstractindextext{%
\begin{abstract}

\vthird{\textit{Kubernetes} (\textit{k8s}) has the potential to coordinate distributed edge resources and centralized cloud resources, but currently lacks a specialized scheduling framework for edge-cloud networks.
Besides, the hierarchical distribution of heterogeneous resources makes} the modeling and scheduling of \textit{k8s}-oriented edge-cloud network particularly \vthird{challenging}.
In this paper, we introduce \textit{KaiS}, a learning-based scheduling framework for such edge-cloud network to improve the long-term throughput rate of request processing.
First, we design a coordinated multi-agent actor-critic algorithm to cater to decentralized request dispatch and dynamic dispatch spaces within the edge cluster.
Second, for diverse system scales and structures, we use graph neural networks to embed system state information, and combine the embedding results with multiple policy networks to reduce the orchestration dimensionality by stepwise scheduling.
Finally, we adopt a two-time-scale scheduling mechanism to harmonize request dispatch and service orchestration, and present the \vthird{implementation} design of deploying the above algorithms compatible with native \textit{k8s} components.
Experiments using real workload traces show that \textit{KaiS} can successfully learn appropriate scheduling policies, irrespective of request arrival patterns and system scales.
Moreover, \textit{KaiS} can enhance the average system throughput rate by $15.9\%$ while reducing scheduling cost by $38.4\%$ compared to baselines.
\end{abstract}

}

\maketitle

\IEEEdisplaynontitleabstractindextext
\IEEEpeerreviewmaketitle

\section{Introduction}
\label{sec:Introduction}

\subsection{Background and Problem Statement}

To provide agile service responses and alleviate \vthird{the burden on backbone networks}, edge and cloud computing are gradually converging to achieve this goal by hosting services as close as possible to where requests are generated \cite{Shi2016}.

\begin{figure}[t]%
	\centering
	\includegraphics[width=8.85 cm]{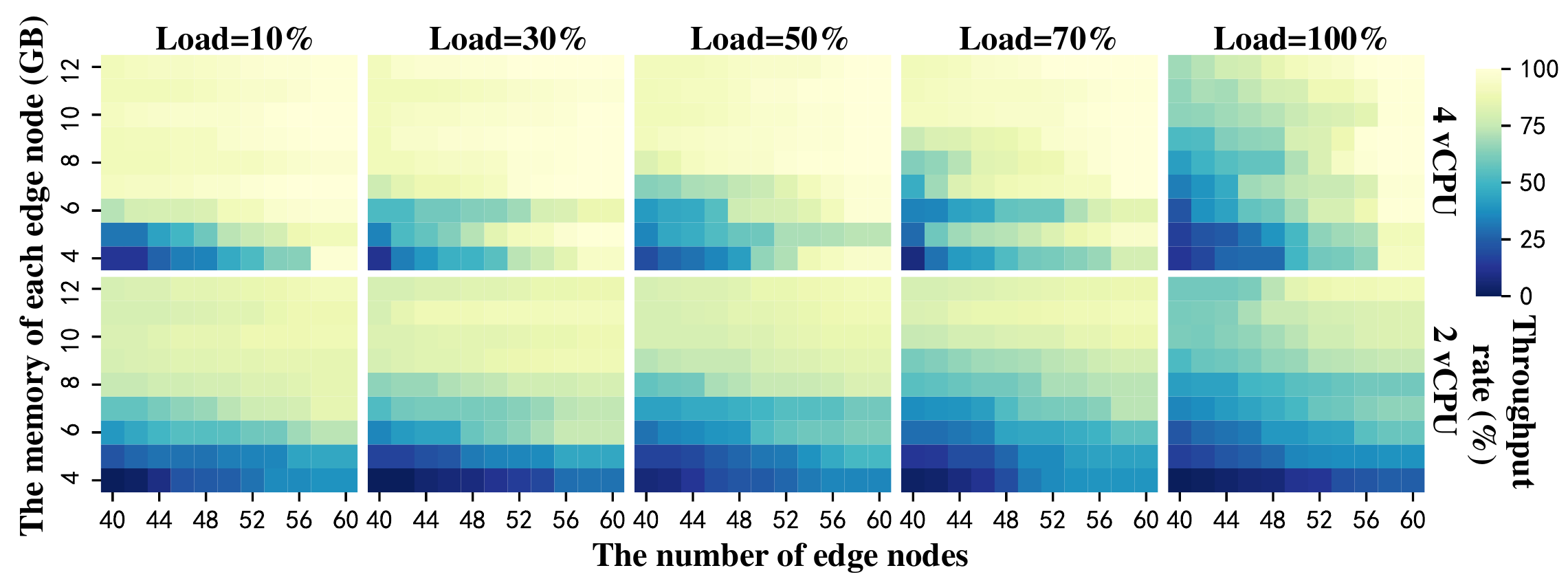}
	\caption{The model of system throughput is highly complex and non-linear.}
	\label{fig:NonlinearEffect}
\end{figure}

Edge-cloud network is commonly built on \textit{Kubernetes} (\textit{k8s}) \cite{Burns2016, kubeedge, openyurt, baetyl} and are designed to seamlessly integrate distributed and hierarchical computing resources at the edge and the cloud \cite{Wang2020}.
In this regard, one fundamental problem for supporting efficient edge-cloud network is: how to schedule request dispatch \cite{Tan2017} and service orchestration (placement) \cite{Pasteris2019} within the \textit{k8s} architecture.
{\color{blue} However, the native \textit{k8s} architecture \vthird{lacks} the ability to manage heterogeneous resources across distributed edge clusters and centralized cloud cluster. In addition, existing customized edge-cloud frameworks (e.g., \textit{KubeEdge} \cite{kubeedge}, \textit{OpenYurt} \cite{openyurt} and \textit{Baetyl} \cite{baetyl}) based on \textit{k8s} do not address the above scheduling issues.}

To serve various requests, the edge-cloud network needs to manage corresponding service entities across \vthird{both edge and cloud while being able to determine} where these requests should be processed. 
Though \textit{k8s} is the most popular tool for managing cloud-deployed services, it is not yet able to accommodate both edge and cloud infrastructure and support request dispatch at the distributed edge. 
\vthird{In this case, the key to an efficient edge-cloud network is to (\textit{$\romannumeral1$}) adapt \textit{k8s} components and extend its current logic to bind the distributed edge and the cloud, and (\textit{$\romannumeral2$}) devise scheduling algorithms that can fit into \textit{k8s}.}
\subsection{Limitations of Prior Art and Motivation}
Most scheduling solutions for request dispatch and service orchestration rely on \vthird{accurately modeling or predicting} service response times, network fluctuation, request arrival patterns\vthird{, and other factors}~\cite{Farhadi2019, Poularakis2019, Ma2020}. 
Nevertheless, (\textit{$\romannumeral1$}) \textit{the heterogeneous edge nodes and the cloud cluster are connected in uncertain network environments, and practically form a dynamic and hierarchical computing system}. 
As shown in Fig.~\ref{fig:NonlinearEffect}, the system behavior, i.e., the average throughput rate of that system managed by native \textit{k8s}, substantially varies with the available resources and the request loads (refer to Sec.~\ref{sec:Performance Evaluation} for detailed settings).
More importantly, (\textit{$\romannumeral2$}) \textit{the underlying model that captures this behavior is highly nonlinear and far from trivial}.
However, even though rich historical data are available, it is hard to achieve the exact estimation of these metrics \cite{Wang2020, Ayala-Romero2019} and then design scheduling policies for any specific request arrivals, system scales and structures, or heterogeneous resources. 
Further, (\textit{$\romannumeral3$}) \textit{few solutions carefully consider whether the proposed scheduling framework or algorithms \vthird{are compatible with} to the actual deployment environment}, i.e., whether they are compatible with \textit{k8s} or others to integrate with the existing cloud infrastructure.
Therefore, a scheduling framework for a \textit{k8s}-oriented edge-cloud network, without relying on the assumption about system dynamics, is desired.

\subsection{Technical Challenges and Solutions}
First, we show the learning-based approach \cite{sutton2018reinforcement, mao2019learning} can improve overall system efficiency by automatically learning effective system scheduling policies \vthird{to cope with stochastic arrivals of service requests.}
We propose \textit{KaiS}, a \textit{\d{k}8s}-oriented and le\d{a}rn\d{i}ng-based scheduling framework for edge-cloud \d{s}ystems.
Given only \vthird{one} high-level goal, e.g., to maximize the long-term throughput of service processing, \textit{KaiS} automatically learns sophisticated scheduling policies \vthird{through the experience of the system operation}, without relying on assumptions about system execution parameters and operating states.

To guide \textit{KaiS} \vthird{in learning} scheduling policies, we need to tailor learning algorithms in the following aspects: \textit{the coordinated learning of multiple agents, the effective encoding of system states, the dimensionality reduction of scheduling actions, etc.}

\begin{figure}[t]%
	\centering
	\includegraphics[width=8.85 cm]{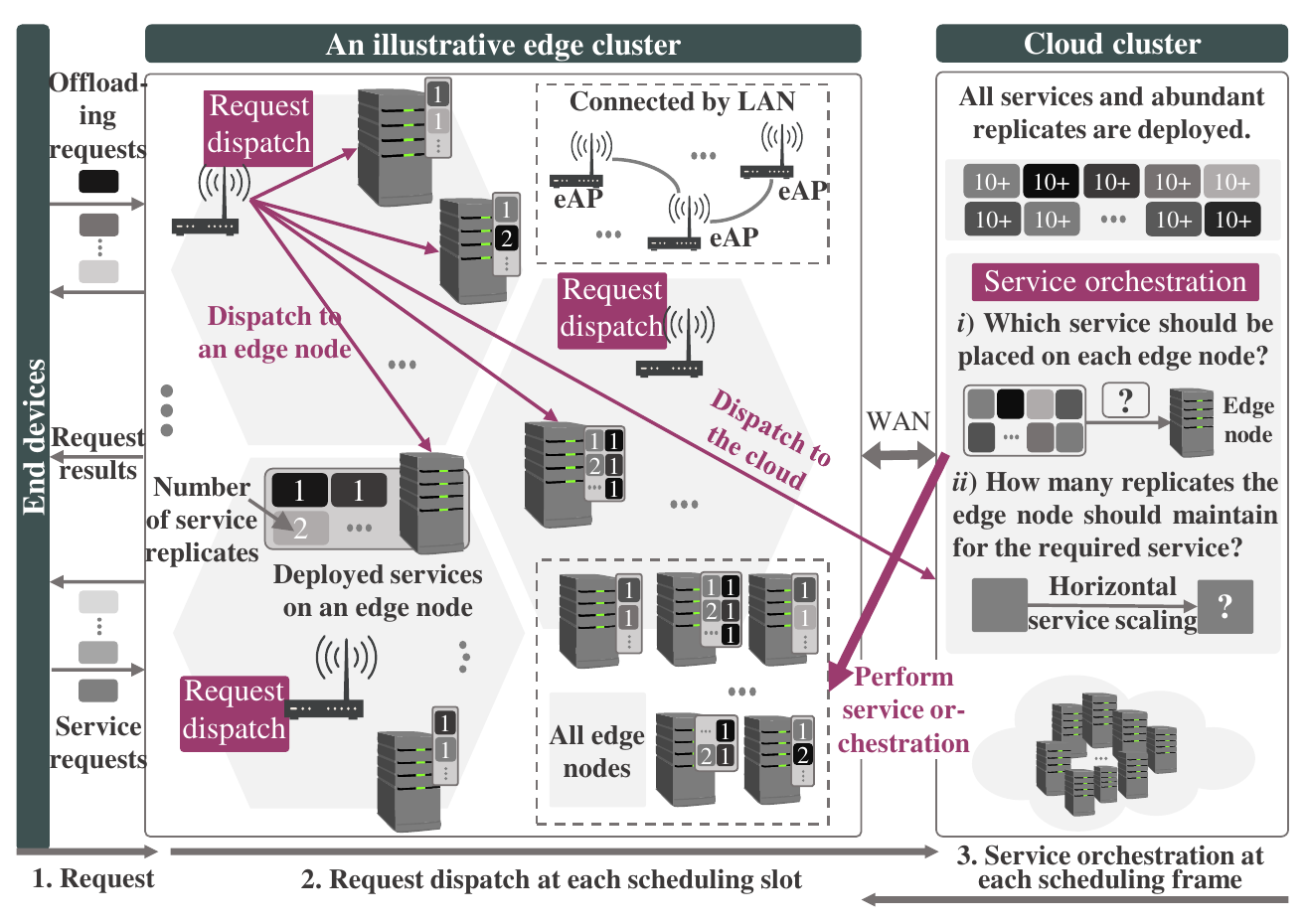}
	\caption{Scheduling in \textit{kubernetes}-oriented edge-cloud network.}
	\label{fig:end-edge-cloud computing}
\end{figure}

For request dispatch, as depicted in Fig.~\ref{fig:end-edge-cloud computing}, \textit{KaiS} needs to scale to hundreds of distributed edge Access Points (eAPs) in the system\cite{Ren2019b}.
Traditional learning algorithms, such as DQN \cite{Mnih2015} and DDPG \cite{lillicrap2015continuous}, that usually use one centralized learning agent, is not feasible for \textit{KaiS} since the distributed eAPs will incur dispatch action space explosion \cite{WangMADRL}.
To ensure timely dispatch, \vthird{\textit{KaiS} requires the dispatch action to be determined in a decentralized manner at the eAPs where the request arrives, rather than using a centralized approach \cite{WangMADRL}.}
Thus, we leverage Multi-Agent Deep Reinforcement Learning (MADRL) \cite{MARLSurvey} and place a dispatch agent at each eAP.

However, such settings (\textit{$\romannumeral1$}) require numerous agents to interact with the system at each time and (\textit{$\romannumeral2$}) have varying dispatch action spaces that depend on available system resources, making these agents difficult to learn scheduling policies.
Hence, we decouple centralized critic and distributed actors, feeding in global observations during critic training to stabilize each agent's learning process, and design a policy context filtering mechanism for actors to respond to the dynamic changes of dispatch action space.

Since different dequeue strategies can affect the performance of request dispatch by influencing the queueing delay of requests, an efficient dequeue strategy is important to optimize system performance.
The request dispatch in the edge-cloud network involves two issues.
The first issue is how to dequeue undispatched requests for dispatch, i.e., after eAPs receive requests from end devices, the requests need to be dequeued for dispatch according to a dequeue strategy. 
The other issue is how to dequeue dispatched requests for processing, i.e., after edge nodes and cloud cluster receive the dispatched requests, the requests need to be dequeued for processing according to a dequeue strategy.
However, requests diversity and environmental dynamics make some classical policies (e.g., first-in-first-out or greedy algorithms) difficult to cope with. 
Therefore, we further consider the characteristics of requests in different request queues and design the dequeue strategy based on discounted experience to calculate the priority of requests.

Besides, \textit{KaiS} must orchestrate dozens of or more services according to the system's global resources and adapt to different system scales and structures.
Hence, \textit{KaiS} requires our learning techniques to (\textit{$\romannumeral1$}) encode massive and diverse edge-cloud system state information, and (\textit{$\romannumeral2$}) represent bigger and complex action space for orchestration in edge-cloud network.
Thus, we employ Graph Neural Networks (GNNs) \cite{Zhang2020} and multiple policy networks \cite{Silver2014} to encode the system information and reduce the orchestration dimensionality, respectively, without manual feature engineering.
Compared with common DRL solutions with raw states and fixed action spaces, our design can reduce model complexity, benefiting the learning of scheduling policies.

\subsection{Main Contributions}
Some of the results of this paper have been presented in the conference version~\cite{kais}. Based on the previous work, this paper extends the work by designing dequeue strategy, refining system design, and adding more experimental results. 

{\color{blue} 
In contrast to state-of-the-art studies, the improvements of \textit{KaiS} mainly consist of the following aspects.
(\textit{$\romannumeral1$}) Most existing studies focus only on either request dispatch, e.g. ~\cite{chen2021multitask, liu2022hastening}, or service orchestration, e.g. ~\cite{lv2022microservice, ding2022kubernetes}, and thus ignore the interaction between services and requests. In contrast, \textit{KaiS} achieves co-optimization of the two aspects through a two-time-scale scheduling framework. (\textit{$\romannumeral2$}) For some algorithm designs based on simulated scenarios, \vthird{they are more theoretical and may have limited practical applicability}, such as considering only a single type of resource limit~\cite{lv2022microservice}, being able to accurately know the amount of computation required for each request~\cite{ wang2021multi}, etc. In contrast, \textit{KaiS} does not \vthird{require assumptions about system state or execution parameters} and can automatically learn appropriate strategies based on experience.  (\textit{$\romannumeral3$}) Many studies focus on theoretical and methodological levels without considering how the algorithms should be deployed in practical application scenarios ~\cite{wang2021multi,jovsilo2020computation}. In contrast, \textit{KaiS} designs non-intrusive components and interaction patterns \vthird{with} k8s (as shown in Sec. IV-B) and uses real-world workload trace from Alibaba~\cite{AlibabaDatasets} for evaluation, guaranteeing applicability in real deployment scenarios.
}
In summary, our main contributions are as follows:

\begin{itemize}[leftmargin=*]
	\item A coordinated multi-agent actor-critic algorithm for decentralized request dispatch with a policy context filtering mechanism that can deal with dynamic dispatch action spaces to address time-varying system resources.
	\item A dequeue strategy based on discounted experience that can adjust the queueing time of requests in the request queue by deciding the request priority, thus reducing the failure of requests due to timeouts.
	\item A GNN-based policy gradient algorithm for service orchestration that employs GNNs to efficiently encode system information and multiple policy networks to reduce orchestration dimensionality by stepwise scheduling.
	\item A two-time-scale scheduling framework implementation of the tailored learning algorithms for the \textit{k8s}-oriented edge-cloud network, i.e., \textit{KaiS}, and an evaluation of \textit{KaiS} with real workload traces (Alibaba Cluster Trace) \cite{AlibabaDatasets} in various scenarios and against baselines.
\end{itemize}

In the sequel, Sec. \ref{sec:Scheduling Problem Statement} introduces the scheduling problem. 
Sec. \ref{sec:Algorithm and System Design} and Sec. \ref{sec:System Design and Implementation} elaborate the algorithm and implementation design. 
Sec. \ref{sec:Performance Evaluation} presents experiment results.
Finally, Sec. \ref{sec:Related Work} reviews related works and Sec. \ref{sec:Conclusion} concludes the paper.

\section{Scheduling Problem Statement}
\label{sec:Scheduling Problem Statement}

We focus on scheduling request dispatch and service orchestration for the edge-cloud network to \textit{improve its long-term throughput rate}, i.e., the ratio of processed requests that meet delay requirements during the long-term system operation. Table \ref{definitions} lists the main notations we will use.

\subsection{Edge-Cloud network}
Edge computing and cloud computing should not be considered mutually exclusive. 
Therefore, to serve the requests from massive end devices distributed at the networks, end-edge-cloud collaborative computing has emerged. 
On the one hand, to accommodate delay-sensitive services, some of these services can be migrated from the cloud computing infrastructure (the cloud) to be deployed in the distributed edge cluster, which consists of multiple neighboring edge nodes. 
On the other hand, when the edge cluster cannot meet the needs of service requests, the cloud can provide powerful backup processing capabilities and high-level global management. 

As shown in Fig.~\ref{fig:end-edge-cloud computing}, neighboring eAPs and edge nodes form a resource pool, i.e., an edge cluster, and connect with the cloud.
When requests arrive at eAPs, the edge cluster then handles them together with the cloud cluster $C$.
For clarity, \textit{we only take one edge cluster to exemplify \textit{KaiS}}, and consider the case that there is no cooperation between geographically distributed edge clusters. {\color{blue}Specifically, each edge cluster is responsible for a geographical area, and there is no collaboration between the edge clusters in different geographical areas.} 
Nonetheless, by maintaining a service orchestrator for each edge cluster, \textit {KaiS} can be easily generalized to support geographically distributed edge clusters.
\begin{itemize}[leftmargin=*]
	\item \textbf{\textit{Edge Cluster and Edge Nodes}}. To process requests, the edge cluster should host corresponding service entities. 
	An edge cluster consists of a set {\color{blue}$\mathcal{B} = \{ {b}_1, {b}_2, \ldots, {b}_B \}$} of eAPs indexed by $b$, and {\color{blue}$\mathcal{N}_{b} = \{ {n}_1, {n}_2, \ldots, {n}_{N_{b}} \}$} is the set of edge nodes attached to and managed by eAP $b$.
	All edge nodes in the edge cluster are represented by {\color{blue}$\mathcal{N}=\sum_{b \in \mathcal{B}} \mathcal{N}_{b}$}.
	All eAPs, along with associated edge nodes, are connected by Local Area Network (LAN).
	A request arrived at the edge can be dispatched to an edge node or the cloud by the eAP that admits it for processing
	.
	\item \textbf{\textit{Cloud Cluster}}. The cloud cluster has sufficient computing and storage resources compared to the edge and is connected to eAPs through Wide Area Network (WAN), 
	It can undertake requests that edge clusters cannot process.
	In addition, it manages all geographically distributed edge clusters, including orchestrating all service entities in each edge cluster according to the system's available resources.
\end{itemize}

\subsection{Improve Long-term System Throughput}
\label{subsec:Scheduling in End-Edge-Cloud Computing}

\begin{table}[t]  
	\centering  
	\begin{threeparttable}  
		\caption{MAIN NOTATIONS}
		\label{definitions}  
		\begin{tabular}{m{1.5cm}<{\centering}m{6.3cm}}  
			\toprule          
			\multicolumn{1}{c}{\bf Notations }&\multicolumn{1}{c}{\bf Descriptions}\cr
			\midrule 

			$a_{b, t}$&The request dispatch action taken by eAP $b$ at slot $t$.\cr
			$b$&The index of eAPs. \cr
			$\mathcal{B}$&The set of eAPs.\cr
			$C$&Cloud cluster.\cr
			$\mathcal{C}_{r_{b, w}}$& The dequeue priority of request $r_{b, w}$.\cr
			$d_{w,n}$&The number of replicates of service $w$ in edge node $n$.\cr
			$\mathcal{E}_{w, t}$& The predicted time to complete a request of type $w$. \cr
			$H$&The number of high-value nodes in each frame.\cr
			$n$&The index of edge nodes.\cr
			$\mathcal{N}$&All edge nodes in the edge cluster.\cr
			$\mathcal{N}_{b}$&The set of edge nodes attached to eAP $b$.\cr
			$\mathcal{Q}_{b}$&Request queue in eAP $b$.\cr
			{\color{blue}$\hat{\mathcal{Q}}_{c}$}&Request queue in cloud cluster.\cr
			{\color{blue}$\mathcal{Q}^{\prime}_{n}$}&Request queue in edge node.\cr
			$r_{b, t}$&The current dispatching request of eAP $b$ at slot $t$.\cr
			{\color{blue}$\mathcal{R}_{c, \tau}$}&{\color{blue}The set of requests received by cloud in frame $\tau$.}\cr
			{\color{blue}$\mathcal{R}_{n, \tau}$}&{\color{blue}The set of requests received by node $n$ in frame $\tau$.}\cr
			
			$t$&The index of slots.\cr
			$\mathcal{T}_{b, w, t}$& The average time consumption corresponding to the request type $w$ calculated at slot $t$ based on the data collected by eAP $b$.\cr
			$w$&The index of services.\cr
			$\mathcal{W}$&The set of all services.\cr
			$\lambda$&The ratio of timeout requests.\cr
			$\lambda_{e}$& The discount factor.\cr
			$\lambda^{\prime}$&The penalty factor.\cr
			$\xi$&The standard deviation of the CPU and memory usage of all edge nodes.\cr
			$\hat{\pi}_{b, t}$&Dynamic request dispatch policy.\cr
			$\tau$&The index of frames.\cr
			$\varUpsilon_{\tau}({\color{blue}\hat{\mathcal{Q}}_{c}})$&The number of requests that have been processed timely by the cloud.\cr
			$\varUpsilon_{\tau}(\mathcal{Q}_{b})$&The number of requests arrived at eAP $b$.\cr
			$\varUpsilon_{\tau}({\color{blue}\mathcal{Q}^{\prime}_{n}})$&The number of requests that have been processed timely by edge node $n$.\cr
			$\varPhi^{\prime}$&The long-term throughput rate.\cr
			$\Psi$& Dequeue strategy for eAPs.\cr
			$\Psi^{\prime}$& Dequeue strategy for edge nodes and cloud cluster.\cr
			\bottomrule  
		\end{tabular}  
	\end{threeparttable}
	
\end{table}

	We are inspired by~\cite{Farhadi2019} to design a two-time-scale mechanism to schedule request dispatch and service orchestration.
	\vthird{Within each time slot $t$, service requests are dispatched to edge nodes or the cloud based on the available resources, while service orchestration is performed in larger time frames $\tau$, which are $\beta$ times longer than a time slot.}
	To improve overall performance, \textit{KaiS} tailors Deep Reinforcement Learning (DRL) to learn scheduling policies through experiences obtained from the running system, and encodes learned policies in neural networks.

\textbf{\textit{Dispatch of Requests at eAPs}}. 
For each eAP $b \in \mathcal{B}$, \vthird{a queue $\mathcal{Q}_{b}$ is maintained for requests. Additionally, a dynamic dispatch policy $\hat{\pi}_{b, t}$ is implemented, which varies with time, along with a dequeue strategy.}
First, delay-sensitive service requests arrive stochastically at queue $\mathcal{Q}_{b}$ and the highest priority request $r_{b, t}$ is dequeued according to the dequeue strategy $\Psi$.
After that, according to $\hat{\pi}_{b, t}$, at each slot $t$, each eAP $b$ dispatches the request $r_{b, t}$ to an edge node, where the required service entity is deployed and that has sufficient resources, or the cloud cluster with sufficient computing resources for processing.
The processing of each request consumes both computation resources and network bandwidth of the edge or the cloud. 
Moreover, dispatching requests to the cloud may lead to extra transmission delay since it is not as close to end devices, i.e., where requests are generated.
Each edge node and the cloud maintain a queue of dispatched requests, i.e., $\{ \mathcal{Q}_n: n \in \mathcal{N} \}$ for edge nodes and ${\color{blue}\hat{\mathcal{Q}}_{c}}$ for the cloud. After that, requests are continuously selected from the request queue $\mathcal{Q}_n$ and ${\color{blue}\hat{\mathcal{Q}}_{c}}$ for processing by the service entities according to dequeue strategy $\Psi^{\prime}$, until no service entity is available to process any request in the queue.
To ensure timely scheduling, it is ideal to have the eAPs, where requests first arrive, perform request dispatch independently, instead of letting the cloud or the edge to make dispatching decisions in a centralized manner, since it may incur high scheduling delays \cite{Chen2016f}.
For requests that are not processed in time, the system drops them at each slot.

\textbf{\textit{Orchestration of Services at Edge Cluster}}. 
{\color{blue}Since each type of request needs to depend on the corresponding type of service entity for processing, placing the service entity on the edge node will occupy the resources of the edge node. 
Therefore, limited by the resource of a single edge node, it is not possible to deploy all types $\mathcal{W} = \{ 1, \ldots, w, \ldots, W \}$ of service entities on each edge node. In addition, due to the resource limitations of a single service entity, it does not have sufficient resources to process when the request load is too high. 
Therefore, multiple replicates of this type of service entity can be placed to share the requests\vthird{, which requires orchestration of the service.}} 
In this case, service entities at the edge cluster should be orchestrated, which includes the following questions: (\textit{$\romannumeral1$}) which service should be placed on which edge node and (\textit{$\romannumeral2$}) how many replicates the edge node should maintain for that service. 
Besides, service requests arrivals at different times may have different patterns, resulting in the intensity of demand for different services varying over time.
Hence, the scheduling should be able to capture and identify such patterns and, based on them, to orchestrate services to fulfill stochastically arriving requests.
Unlike request dispatch, too frequent large-scale service orchestration in the edge cluster may incur system instability and high operational costs due to resource constraints\cite{Farhadi2019}.
For these reasons, a more appropriate solution is to have the cloud perform service orchestration for the edge with a dynamic scheduling policy $\tilde{\pi}_{\tau}$ at each frame $\tau$.
\vthird{Therefore, the cloud can perform service orchestration for the edge with a dynamic scheduling policy $\tilde{\pi}_{\tau}$ at each frame $\tau$, which determines the number of replicates of service $w$ on edge node $n$, denoted by $d_{w,n} \in \mathbb{N}$. If $d_{w,n} = 0$, it means that edge node $n$ does not host service $w$.}

The scheduling objective is to maximize the long-term system throughput $\varPhi = {\color{blue}\sum\nolimits_{\tau=0}^{\infty}}\sum\nolimits_{n \in \mathcal{N}} \varUpsilon_{\tau}({\color{blue}\mathcal{Q}^{\prime}_{n}}) +\varUpsilon_{\tau}({\color{blue}\hat{\mathcal{Q}}_{c}})$, where $\varUpsilon_{\tau}({\color{blue}\mathcal{Q}^{\prime}_{n}})$, $\varUpsilon_{\tau}({\color{blue}\hat{\mathcal{Q}}_{c}})$ represent the number of requests that have been processed timely by edge node $n$ or the cloud in frame $\tau$, respectively. {\color{blue} Specifically, we first denote the set of requests received by node $n$ and the cloud in frame $\tau$ as $\mathcal{R}_{n, \tau}$ and $\mathcal{R}_{c, \tau}$ respectively, then for each request $r \in \mathcal{R}_{n, \tau} $ there will be a delay requirement $t_r$ and an actual completion time $t_r^{\prime}$. Further, we define that if $t_r^{\prime} \leq t_r$ then $\Gamma(r)=1$, otherwise,  $\Gamma(r)=0$. Finally, we obtain $\varUpsilon_{\tau}({\color{blue}\mathcal{Q}^{\prime}_{n}})=\sum_{r \in \mathcal{R}_{n, \tau}} \Gamma(r)$, $\varUpsilon_{\tau}({\color{blue}\mathcal{Q}^{\prime}_{c}})=\sum_{r \in \mathcal{R}_{c, \tau}} \Gamma(r)$.}
To avoid $\varPhi \rightarrow \infty$, we use a more realistic metric, i.e., the long-term system throughput rate $\varPhi^{\prime} \in [0,1]$, which is the ratio of requests, completed within delay requirements, to the total number of requests.
The long-term throughput rate $\varPhi^{\prime}$ can be denoted as $\varPhi^{\prime} = \varPhi / {\color{blue}\sum\nolimits_{\tau=0}^{\infty}}\sum\nolimits_{b \in \mathcal{B}} \varUpsilon_{\tau}(\mathcal{Q}_{b}) $, where $\varUpsilon_{\tau}(\mathcal{Q}_{b})$ indicates the number of requests arrived at eAP $b$ during frame $\tau$.
In this case, our scheduling problem for both request dispatch and service orchestration can be formulated as
{\color{blue}
\begin{equation}
	\label{equ:OptimizationProblem}
	\underset{  \{\hat{\pi}_{b, t} : b \in \mathcal{B} \}, \tilde{\pi}_{\tau} }{\max} \varPhi^{\prime} = \underset{  \{\hat{\pi}_{b, t} : b \in \mathcal{B} \}, \tilde{\pi}_{\tau} }{\max} \varPhi / \sum_{\tau=0}^{\infty}\sum_{b \in \mathcal{B}} \varUpsilon_{\tau}(\mathcal{Q}_{b}) ,
\end{equation}
}
where, for clarity, we use scheduling policies $\{ \hat{\pi}_{b, t}:b \in \mathcal{B} \}$ and $\tilde{\pi}_{\tau}$ instead of a series of scheduling variables to represent the problem.
Compared to the problem in \cite{Farhadi2019}, our scheduling is more complicated since it involves integer dispatch variables.
More details on the constraints and NP-hard proof of such a long-term scheduling problem can be found in \cite{Farhadi2019}.
In this work, we tailor learning algorithms for \textit{KaiS} to improve the long-term system throughput rate.

\vspace{-0.3em}
\begin{algorithm}
	\caption{\small Training and Scheduling Process of \textit{KaiS}}
	\label{algorithm:global}
	\small Initialize the system environment and neural networks.

	\For{\rm{slot} $t$ = $1, 2, ...$}{

		\If {\rm{frame} $\tau$ \rm{begins}}{
			
			Get reward $\tilde{u}_{\tau-1}$ and store $[\tilde{\boldsymbol{s}}_{\tau-1}, \tilde{\boldsymbol{a}}_{\tau-1}, \tilde{u}_{\tau-1}] $;
			
			Use GNNs to embed system states as Eq. (\ref{equ:gnn embedding});
			
			Select $H$ high-value edge nodes $(\tilde{a}_{\tau}^{\bullet})$ and
			\linebreak compute their service scaling actions $(\tilde{\boldsymbol{a}}_{\tau}^{\star})$  \hspace*{1.8em}%
			\hbox{\raise 5pt \rlap{\smash{$\left.\begin{array}{@{  }c@{}}\\{}\\{}\\{}\\{}\\{}\\{}\\{}\end{array}\color{orchestrator2}\right\}$}}}
			\hbox{\raise 9pt \rlap{\smash{$\color{orchestrator2}\begin{tabular}{l}\rotatebox{270}{\footnotesize \textbf{\textit{Orchestrate (GPG)}}}\end{tabular}$}}} 
			\linebreak using policy networks $\theta_g$ and $\theta_q$, respectively;

			Execute orchestration action $\tilde{\boldsymbol{a}}_{\tau} = (\tilde{a}_{\tau}^{\bullet}, \tilde{\boldsymbol{a}}_{\tau}^{\star})$;
			
			Update GNNs and policy networks by Eq. (\ref{equ:GNNDRL policy gradient});
		}
		
		 \ForPar{\rm{each eAP agent} $b \in B$ }{
			
			\If{ $\mathcal{Q}_{b} == \varnothing$}{
			
				\textbf{Continue}
			}
			
			Update request queue $\mathcal{Q}_{b}$ and get reward $\hat{u}_{b, t-1}$;
			
			Store $[ \hat{\boldsymbol{s}}_{b, t-1}, \hat{a}_{b, t-1}, A\left(\hat{\boldsymbol{s}}_{b, t-1}, \hat{a}_{b, t-1}\right),\hat{u}_{b, t-1},$
			\linebreak $ \boldsymbol{F}_{b, t-1}] $ for $\theta_{p}$ (actor);
			
			 Dequeue request $r_{b, t}$ according to $\Psi$; 
			
			Compute the resource context $\boldsymbol{F}_{b, t}$ using Eq. (\ref{equ:resource context});
			
			Take dispatch action $\hat{a}_{b, t}$ for $r_{b,t}$ using Eq. (\ref{equ:probability of valid actions for agent});\hspace*{1.1em}%
			\hbox{\raise 0pt \rlap{\smash{$\left.\begin{array}{@{}c@{}}\\{}\\{}\\{}\\{}\\{}\\{}\\{}\\{}\\{}\\{}\end{array}\color{dispatchers2}\right\}$}}}
			\hbox{\raise -0pt \rlap{\smash{$\color{dispatchers2}\begin{tabular}{l}\rotatebox{270}{\textbf{\footnotesize \textit{Dispatch (cMMAC)}}}\end{tabular}$}}}
			
		}
	    Execute dispatched requests according to $\Psi^{\prime}$;
		
		Store $[ \hat{\boldsymbol{s}}_{t-1}, V^{*}\left(\hat{\boldsymbol{s}}_{t} ; \theta_{v}^{\prime}, \pi\right)] $ for $\theta_{v}$ (critic);
		
		Update neural networks $\theta_{p}$ (actor) and $\theta_{v}$ (critic) 
		\linebreak centrally using Eq. (\ref{equ:the gradient of policy of cMAAC}) and  (\ref{equ:loss function derived from Bellman equation}), respectively; 
		
		Synchronize $\theta_{p}$ periodically to distributed eAPs.

	} 
\end{algorithm}

\section{Algorithm Design}
\label{sec:Algorithm and System Design}

The overall training and scheduling process of \textit{KaiS} is given in Algorithm \ref{algorithm:global}. 
We explain the technical details of request dispatch and service orchestration in the following.
Detailed training settings are presented in Sec. \ref{subsec:Training Settings}.

\subsection{Tailored MADRL for Decentralized Request Dispatch}

Request dispatch is to let each eAP independently decide which edge node or the cloud should serve the arrived request.
The goal of dispatch is to maximize the long-term system throughput rate $\varPhi^{\prime}$ by (\textit{$\romannumeral1$}) balancing the workloads among edge nodes and (\textit{$\romannumeral2$}) further offloading some requests to the cloud in some appropriate cases.

\subsubsection{Markov Game Formulation}

To employ MADRL, we formulate that eAPs independently perform request dispatch as a Markov game $\mathcal{G}$ for eAP agents $\mathcal{B} = \{ 1, 2, \ldots, B \}$.
Formally, the game $\mathcal{G} = ( \mathcal{B}, \hat{\mathcal{S}}, \hat{\mathcal{A}}, \hat{\mathcal{P}}, \hat{\mathcal{U}} )$ is defined as follows.
\begin{itemize}[leftmargin=*]
	\item \textbf{\textit{State}}. 
	$\hat{\mathcal{S}}$ is the state space.
	At each slot $t$, we periodically construct a local state $\hat{\boldsymbol{s}}_{b, t}$ for each eAP agent $b$, which consists of (\textit{$\romannumeral1$}) the service type and delay requirement of the current dispatching request $r_{b, t}$, (\textit{$\romannumeral2$}) the queue information $\mathcal{Q}_{b, t}^{\prime}$ of requests awaiting dispatch at eAP $b$, (\textit{$\romannumeral3$}) the queue information, $\{ \mathcal{Q}_{n_b, t}^{\prime}:n_b \in \mathcal{N}_b \}$, of unprocessed requests at edge nodes $\mathcal{N}_b$, (\textit{$\romannumeral4$}) the remaining CPU, memory and storage resources of $\mathcal{N}_b$, (\textit{$\romannumeral5$}) the number of $\mathcal{N}_b$, i.e., $|\mathcal{N}_b| = N_{b}$, and (\textit{$\romannumeral6$}) the measured network latency between the eAP and the cloud.
	Meanwhile, for centralized critic training, we maintain a global state $\hat{\boldsymbol{s}}_{t} \in \hat{\mathcal{S}}$, which includes (\textit{$\romannumeral1$}) the above information for all eAPs $\mathcal{B}$ and edge nodes $\mathcal{N}$, instead of only eAP $b$ and $\mathcal{N}_b$, and (\textit{$\romannumeral2$}) the queue information $\mathcal{Q}_{C, t}^{\prime}$ of unprocessed requests at the cloud cluster $C$.

	\item \textbf{\textit{Action space}}. The joint action space of $\mathcal{G}$ is $\hat{\mathcal{A}}=\hat{\mathcal{A}}_{1} \times \ldots \times \hat{\mathcal{A}}_{b} \times \ldots \times \hat{\mathcal{A}}_{B}$. %
	For an edge cluster, we consider all available edge nodes as a resource pool, namely the cooperation between eAPs in enabled. In this case, $\hat{\mathcal{A}}_{b}$ includes $N+1$ discrete actions denoted by ${\{i\}}_{0}^{N}$, where $a_{b, t} = 0$ and $a_{b, t} \in \mathcal{N}$ specify dispatching to the cloud or edge nodes, respectively. At each slot $t$, we use $ \hat{\boldsymbol{a}}_{t} = ( a_{b, t}: b \in \mathcal{B} ) $ to represent the joint dispatch actions of all requests required to be scheduled at all eAPs. Note that multiple requests may be queued in eAPs ($\mathcal{Q}_{b}, b \in \mathcal{B}$), we only allow each eAP agent to dispatch one request at a slot. Meanwhile, for \textit{KaiS}, we set the time slot to a moderate value (refer to Sec. \ref{sec:System Design and Implementation}) to ensure the timeliness of serving request arrivals.
	\item \textbf{\textit{Reward function}}. All agents in the same edge cluster share a reward function $\hat{U} = \hat{\mathcal{S}} \times \hat{\mathcal{A}} \rightarrow \mathbb{R} $, i.e., $\hat{U}_b = \hat{U}$ holds for all $b \in \mathcal{B}$. Each agent wants to maximize its own expected return $\mathbb{E}\left[\sum\nolimits_{i=0}^{\infty} \gamma^{i} \hat{u}_{b, t+i}\right]$, where $\hat{u}_{b, t}$ is the immediate reward for the $b$-th agent associated with the action $a_{b, t}$ and $\gamma \in (0,1]$ is a discount factor. The immediate reward is defined as $\hat{u}_{b, t} = {\text{e}}^{-\lambda - \varepsilon \nu }$. Specifically, (\textit{$\romannumeral1$}) $\lambda \in [0, 1]$ is the ratio of requests that violate delay requirements during $[t, t+1]$, (\textit{$\romannumeral2$}) $\nu = 1 / (1+\mathrm{e}^{-\xi}) \in [0.5, 1]$, where $\xi \in \mathbb{R}_{\ge 0}$ is the standard deviation of the CPU and memory usage of all edge nodes, and (\textit{$\romannumeral3$}) $\varepsilon$ is the weight to control the degree of load balancing among edge nodes. The introduction of $\nu$ is to stabilize the system, preventing too much load are imposed on some edge nodes. When $\nu$ is closer to $0.5$, i.e., $\xi \to 0$, the loads of edge nodes are more balanced, thus leading to more scheduling rooms for dispatch. Such a reward is to \textit{improve the long-term throughput while ensuring the load balancing at the edge}. 
	\item \textbf{\textit{State transition probability}}. We use $ p \left( \hat{\boldsymbol{s}}_{t+1} \mid \hat{\boldsymbol{s}}_{t}, \hat{\boldsymbol{a}}_{t} \right): \hat{\mathcal{S}} \times \hat{\mathcal{A}} \times \hat{\mathcal{S}} \rightarrow [0, 1] $ to indicate the transition probability from state $\hat{\boldsymbol{s}}_{t}$ to $\hat{\boldsymbol{s}}_{t+1}$ given a joint dispatch action $\hat{\boldsymbol{a}}_{t}$. The action is deterministic in $\mathcal{G}$, i.e., if $a_{b, t} = 2$, the agent $b$ will dispatch the current request to edge node $2$ at slot $t+1$. 
\end{itemize}

\subsubsection{Coordinated Multi-Agent Actor-Critic}
\label{subsubsec:Coordinated Multi-agent Actor-Critic}

\begin{figure}[t]%
	\centering
	\includegraphics[width=8.55 cm]{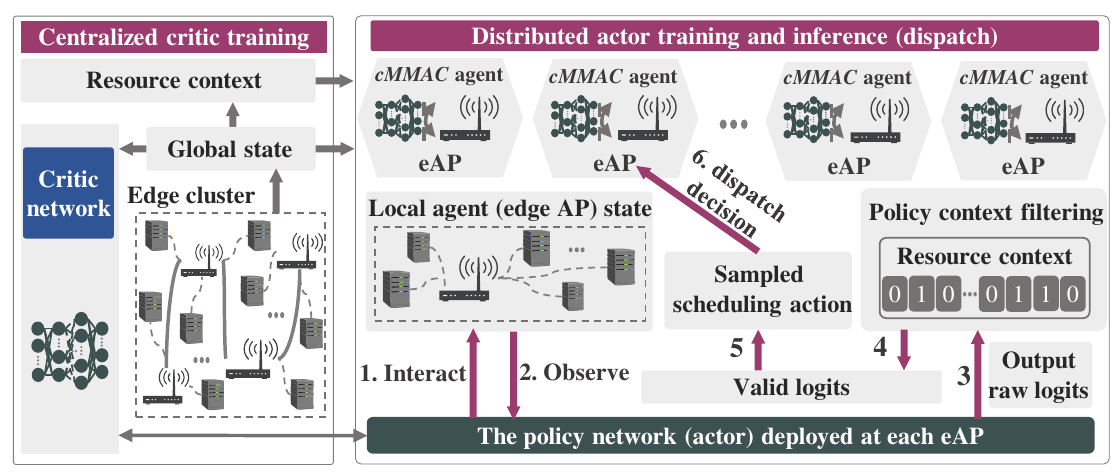}
	\caption{Coordinated multi-agent actor-critic for request dispatch.}
	\label{fig:cMMAC}
\end{figure}

The challenges of training these dispatch agents are:
(\textit{$\romannumeral1$}) The environment of each agent is non-stationary \vthird{as} other agents are learning and affecting the environment simultaneously. 
Specifically, each agent usually learns its own policy that is changing over time \cite{MARLSurvey}, \vthird{which increases the difficulty of coordination;}
(\textit{$\romannumeral2$}) The action space of each agent changes dynamically since its feasible dispatch options vary with the available system resources, making vanilla DRL algorithms unable to handle. 
{\color{blue}For instance, if the memory of the edge node $n$ is run out at slot $t$, the action space should not include the option of dispatching the request to edge node $n$.}

Therefore, we design \textit{coordinated Multi-Agent Actor-Critic (cMMAC)}, as illustrated in Fig.~\ref{fig:cMMAC}: 
(\textit{$\romannumeral1$}) Adopt a centralized critic and distributed actors to coordinate learning, i.e., all agents share a centralized state-value function when training critic, while during distributed actor training and inference each actor only observes the local state. 
(\textit{$\romannumeral2$}) By policy context filtering, we can adjust their policies to tolerate dynamic action space and establish explicit coordination among agents to facilitate successful training.
The details illustrated as follows.

\begin{itemize}[leftmargin=*]
	\item \textbf{\textit{Centralized state-value function (Critic)}}. The state-value function shared by eAP agents can be obtained by minimizing the loss function derived from Bellman equation \cite{sutton2018reinforcement}, which is as follows:
	{\color{blue}
	\begin{equation}
		\label{equ:loss function derived from Bellman equation}
		L\left(\theta_{v}\right)=\left(V_{\theta_{v}}\left(\hat{\boldsymbol{s}}_{t-1}\right)-V^{\ast}\left(\hat{\boldsymbol{s}}_{t} ; \theta_{v}^{\prime}, \pi\right)\right)^{2},
	\end{equation}
	\begin{multline}
		\label{equ:centralized state-value function}
		V^{\ast}\left(\hat{\boldsymbol{s}}_{t}; \theta_{v}^{\prime}, \pi\right)= \\ \sum\nolimits_{a_{b, t-1} \in \hat{\boldsymbol{a}}_{t-1}} \pi\left(a_{b, t-1} \mid \hat{\boldsymbol{s}}_{b, t-1}\right)\left(\hat{u}_{b, t-1}+\gamma V_{\theta_{v}^{\prime}}\left(\hat{\boldsymbol{s}}_{b, t}\right)\right),
	\end{multline}
}
	where $\theta_{v}$ and $\theta_{v}^{\prime}$ denote the parameters of the value network and the target value network, {\color{blue}and $\pi\left(a_{b, t-1} \mid \hat{\boldsymbol{s}}_{b, t-1}\right)$ denotes the probability of adopting $a_{b, t-1}$ at $\hat{\boldsymbol{s}}_{b, t-1}$.}
	In total, for $B$ eAP agents, there are $B$ unique state-values $\{ V(\hat{\boldsymbol{s}}_{b, t-1}): b \in \mathcal{B} \}$ at each slot. 
	Each state-value output $V(\hat{\boldsymbol{s}}_{b, t-1})$ is the expected return received by agent $b$ at slot $t$.
	To stabilize the state-value function, we fix a target value network $V^{\ast}$ parameterized by $\theta_{v}^{\prime}$ and update it at the end of each training episode.
	\item \textbf{\textit{Policy context filtering (Actors)}}. 
	{\color{blue}Due to the neural network structure, the action space for actors has a fixed size. However, the size of the actual available action space changes dynamically during the operation of the system. To solve this problem, we design the policy context filtering for actors. Specifically, during the operation of the edge-cloud network, randomly arrived requests cause the available resources of edge nodes to change dynamically. Therefore, if the available resources of an edge node are insufficient to process the current request, dispatching the request to that node is an invalid action. Thus, to avoid such invalid actions as much as possible, before dispatching, we compute a resource context $\boldsymbol{F}_{b, t} \in \{0, 1\}^{N+1}$ for each eAP agent, which is a binary vector that filters out invalid dispatch actions.}
	 In detail, the value of the element of $\boldsymbol{F}_{b, t}$ is defined as:
	\begin{equation}
		\label{equ:resource context}
		\left[\boldsymbol{F}_{b, t}\right]_{j}=\left\{\begin{array}{ll}
			1, & \text{ if edge node } j \text{ is available,} \\
			1, & \text{ if } j = 0, \\
			0, & \text { otherwise. }
		\end{array}\right.
	\end{equation}
	where (\textit{$\romannumeral1$}) $[\boldsymbol{F}_{b, t}]_{j}$ ($j = 1, \ldots, N$) represents the validity of dispatching the current request to $j$-th edge node and (\textit{$\romannumeral2$}) $[\boldsymbol{F}_{b, t}]_{0}$ specifies that the cloud cluster ($j=0$) is always a valid action of request dispatch, namely $\left[\boldsymbol{F}_{b, t}\right]_{0} \equiv 1$.
	In addition, the coordination of agents is also achieved by masking available action space based on the resource context $\boldsymbol{F}_{b, t}$.
	To proceed, we first use $\boldsymbol{p}(\hat{\boldsymbol{s}}_{b,t}) \in \mathbb{R}^{N+1}$ to denote the original output logits from the actor policy network for the $b$-th agent conditioned on state $\hat{\boldsymbol{s}}_{b,t}$. Next, we let $\bar{\boldsymbol{p}}(\hat{\boldsymbol{s}}_{b, t})=\boldsymbol{p}(\hat{\boldsymbol{s}}_{b, t}) * \boldsymbol{F}_{b, t}$, where the operation $*$ is element-wise multiplication, to denote the valid logits considering the resource context for agent $b$. Besides, note that the output logits $\boldsymbol{p}(\hat{\boldsymbol{s}}_{b,t}) \in \mathbb{R}^{N+1}_{>0}$ are restricted to be positive to achieve effective masking. Hence, the probability of valid dispatch actions for agent $b$ can be given by:
	\begin{equation}
		\label{equ:probability of valid actions for agent}
		\pi_{\theta_{p}}\left(a_{b,t}=j \mid \hat{\boldsymbol{s}}_{b,t}\right)=\left[\bar{\boldsymbol{p}}\left(\hat{\boldsymbol{s}}_{b,t}\right)\right]_{j}=\frac{[\bar{\boldsymbol{p}}(\hat{\boldsymbol{s}}_{b,t})]_{j}}{\|\bar{\boldsymbol{p}}(\hat{\boldsymbol{s}}_{b, t})\|_{1}}, 
	\end{equation}
	where $\theta_{p}$ is the parameters of actor policy network. At last, the policy gradient $\nabla_{\theta_{p}} J(\theta_{p})$ can be derived and the advantage $A(\hat{\boldsymbol{s}}_{b, t}, a_{b, t})$ {\color{blue}(related to the use of the target network parameters in~\cite{mnih2015human})} can be computed.
	\begin{equation}
		\label{equ:the gradient of policy of cMAAC}
		\nabla_{\theta_{p}} J\left(\theta_{p}\right)=\nabla_{\theta_{p}} \log \pi_{\theta_{p}}\left(a_{b, t} \mid \hat{\boldsymbol{s}}_{b, t}\right) A\left(\hat{\boldsymbol{s}}_{b, t}, a_{b, t}\right),
	\end{equation}
	\begin{equation}
		\label{equ:the advantage of cMAAC}
		A\left(\hat{\boldsymbol{s}}_{b, t}, a_{b, t}\right)=\hat{u}_{b, t+1}+\gamma V_{\theta_{v}^{\prime}}\left(\hat{\boldsymbol{s}}_{b, t+1}\right)-V_{\theta_{v}}\left(\hat{\boldsymbol{s}}_{b, t}\right).
	\end{equation}
	
\end{itemize}

\subsubsection{Dequeue Strategy for Request Priority}
\label{subsubsec:Dequeue_Strategy_for_Request_Priority}
\begin{figure}[t]%
	\centering
	\includegraphics[width=8.95 cm]{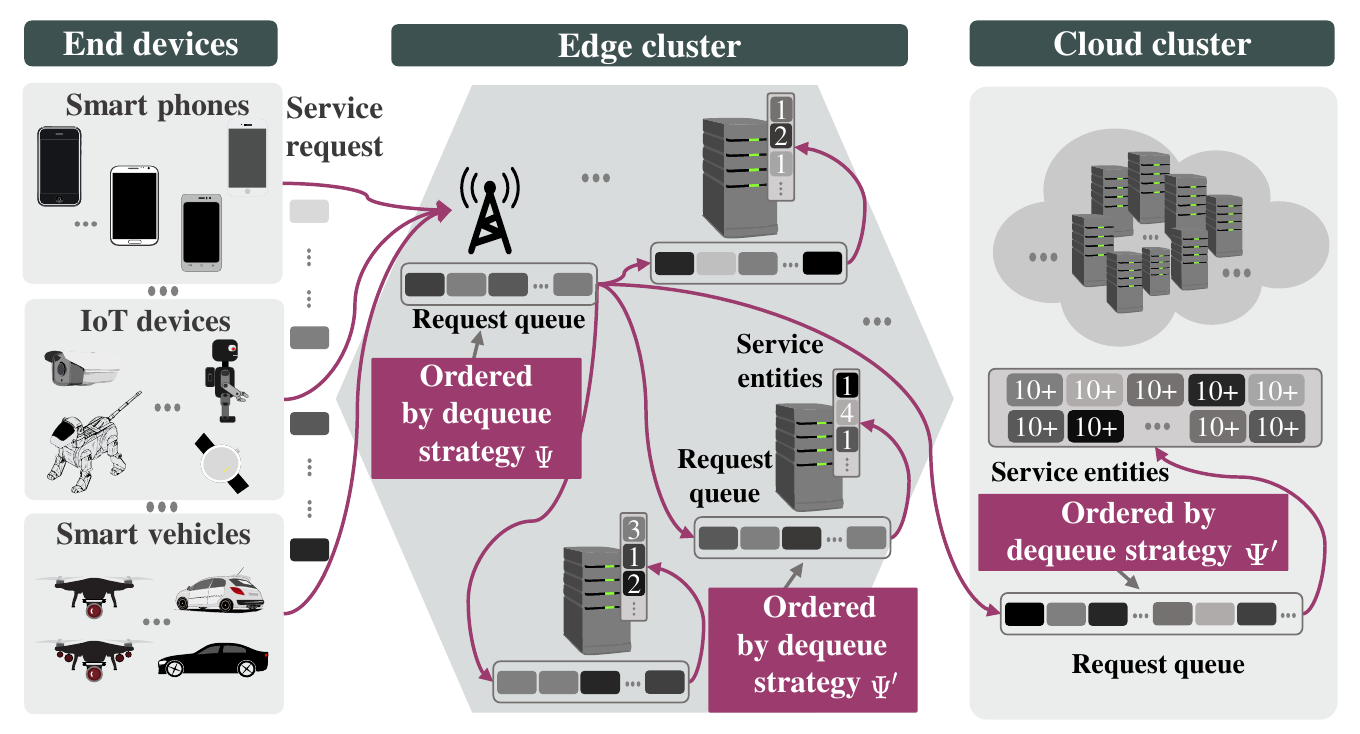}
	\caption{Dequeue strategies in edge cluster and cloud cluster.}
	\label{fig:Dequeue_Fig}
\end{figure}

For request queues, \vthird{two issues that need to be considered are}: (\textit{$\romannumeral1$}) which request is selected to be dispatched after the eAPs receive the randomly arriving requests; (\textit{$\romannumeral2$}) how to decide the order of processing these requests after the edge nodes and cloud cluster receive the dispatched requests. As shown in Fig.~\ref{fig:Dequeue_Fig}, we design a dequeue strategy as follows.

\begin{itemize}[leftmargin=*]
\item \textbf{\textit{Dequeue strategy $\Psi$ in eAPs}}. 
For each service type $w \in \mathcal{W}$, \vthird{we define the estimated time} to complete a request of type $w$ as $\mathcal{E}_{w, t}$. To accommodate environment dynamics, $ \mathcal{E}_{w, t}$ can be updated dynamically and a discount factor $ \lambda_{e} \in (0, 1)$ is designed to give higher weight to the recent experience. 
\vthird{In particular, the experience here refers to the historical data generated in the system}, such as the transmission time, execution time, and queueing time of each completed request.
To update $\mathcal{E}_{w, t}$ at slot $t$, eAP $b$ first collects the actual consumption time of requests from dequeue to completion.
Then, the collected data is categorized by eAP $b$ according to request type to calculate the actual average time consumption $  \mathcal{T}_{b, w, t}$ for request type $w$.
If eAP $b$ does not collect any request consumption time for service type $w$ at slot $t$, then $  \mathcal{T}_{b, w, t}$ takes the value of zero. Finally, for the actual average time consumption $  \mathcal{T}_{b, w, t} \neq 0$, the algorithm updates $ \mathcal{E}_{w, t}$ by:

\begin{equation}
	\label{equ:expected_completion_time}
	\mathcal{E}_{w, t+1} = \lambda_{e} \mathcal{E}_{w, t} + (1-\lambda_{e}) \mathcal{T}_{b, w, t}.
\end{equation}

After that, for each request $r_{b, w}$ in the request queue of eAP $b$, we define the remaining time from its deadline as $\mathcal{F}_{r_{b, w}}$, which can be calculated based on request generation time, delay limit and current time.
Furthermore, it is more likely that the request of $  \mathcal{F}_{r_{b, w}} \textless \mathcal{E}_{w, t}$ cannot be completed on time, i.e., the remaining time before the deadline is less than the predicted time required. Therefore, we add a penalty factor $\lambda^{\prime} \in (0, 1)$ in the calculation of the priority $\mathcal{C}_{r_{b, w}}$ as follows:

\begin{equation}
	\label{equ:priority}
	\mathcal{C}_{r_{b, w}}=\left\{\begin{array}{ll}
	1/( \mathcal{F}_{r_{b, w}}-\mathcal{E}_{w, t}), & \text{ if } \mathcal{F}_{r_{b, w}} \textgreater \mathcal{E}_{w, t}, \\
	\lambda^{\prime}/(\mathcal{E}_{w, t}- \mathcal{F}_{r_{b, w}}), & \text{ if }  \mathcal{F}_{r_{b, w}} \textless \mathcal{E}_{w, t}\text{.}
	\end{array}\right.
\end{equation}

Finally, the requests in the request queue are sorted according to priority $\mathcal{C}_{r_{b, w}}$ , and the requests with the highest priority are dequeued for scheduling.

\item \textbf{\textit{Dequeue strategy $\Psi^{\prime}$ on edge nodes and cloud cluster}}. 
Similar to the dequeue strategy $\Psi$ at eAPs, the priority $\mathcal{C}^{\prime}_{r_{b, w}}$ is computed based on (\ref{equ:expected_completion_time}) and (\ref{equ:priority}) to determine the order in which the requests are dequeued from the queue. Note that there are two adjustments that differ from the dequeue strategy $\Psi$ in eAPs:
(\textit{$\romannumeral1$}) the transmission time of requests is ignored compared to the dequeue strategy $\Psi$ in eAPs, i.e., the dequeue strategy $\Psi^{\prime}$ in each edge node or cloud cluster only considers the time consumption after requests are dequeued from the local request queue;
 (\textit{$\romannumeral2$}) the number of requests dequeued in a slot is not one, compared to the dequeue strategy $\Psi$ in eAPs, i.e., each request in the request queue is checked from the highest to the lowest priority to see if any service entity can match it, and if there is a corresponding service entity with free resources, the request is dequeued and the next lower priority request is checked.
\end{itemize}

\subsection{GNN-based Learning for Service Orchestration}
\label{subsec:GNN-based DRL for Service Orchestration}

In this section, we propose a \textit{GNN-based Policy Gradient (GPG)} algorithm and describe how 
(\textit{$\romannumeral1$}) the system state information is processed flexibly;
(\textit{$\romannumeral2$}) the high-dimensional service orchestration is decomposed as stepwise scheduling actions: selecting high-value edge nodes and then performing service scaling on them.

For edge clusters with different scales, composed of edge nodes with different numbers and heterogeneous resources, we use the GNNs described in Sec. \ref{subsec:Training Settings}, namely parametrized functions learned during training, to repeatedly process the information of the edge-cloud network concerning the network topologies, real-time available resources, service request queues, etc. 
In addition, the large decision interval of service orchestration and the centralized decision making approach pose a challenge to the collection of training data samples. The multiple neural networks in the actor-critic algorithm and large number of neural network parameters would further increase the difficulty of training.
Therefore, for better convergence, \vthird{we propose the policy gradient algorithm to solve the service orchestration problem.}

\subsubsection{GNN-based System State Encoding}
\label{subsubsec:GNN-based System State Encoding}

\begin{figure}[t]%
	\centering
	\includegraphics[width=8.85 cm]{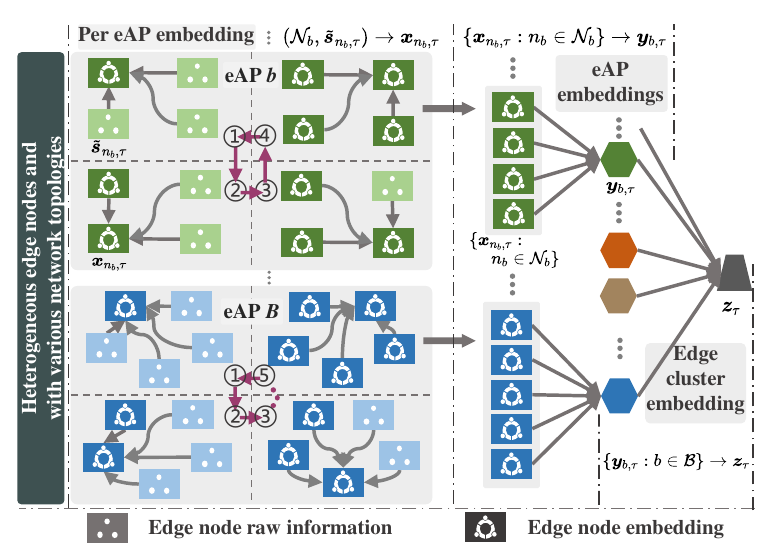}
	\caption{GNN-based system state encoding for the edge cluster.}
	\label{fig:GNNEncode} 
\end{figure}

As shown in Fig.~\ref{fig:GNNEncode}, \textit{KaiS} must convert system states into feature vectors on each observation and then pass them to policy networks.
A common choice is directly stacking system states into flat vectors.
However, the edge-cloud network is practically a graph consisting of connected eAPs, edge nodes, and the cloud cluster.
Simply stacking states has two defects: 
(\textit{$\romannumeral1$}) processing a high-dimensional feature vector requires sophisticated policy networks, which increases training difficulty; 
(\textit{$\romannumeral2$}) it cannot efficiently model the graph structure information for the system, making \textit{KaiS} hard to generalize to various system scales and structures.
Therefore, we use GNNs to encode system states into a set of embeddings layer by layer as follows.
\begin{itemize}[leftmargin=*]
	\item \textbf{\textit{Embedding of edge nodes}}.
	For edge nodes associated with eAP $b$, each of them, $n_b \in \mathcal{N}_b$, carries the following attributes at each frame $\tau$, denoted by a vector $\tilde{\boldsymbol{s}}_{n_b, \tau}$: (\textit{$\romannumeral1$}) the available resources of CPU, memory, storage, etc., (\textit{$\romannumeral2$}) the periodically measured network latency with eAP $b$ and the cloud, (\textit{$\romannumeral3$}) the queue information of \vthird{backlogged requests it is currently processing}, i.e., $\mathcal{Q}_{n_b}^{\prime}$, and (\textit{$\romannumeral4$}) the indexes of deployed services and the number of replicates of each deployed service. 
	Given $\tilde{\boldsymbol{s}}_{n_b, \tau}$, \textit{KaiS} performs embedding for each edge node as $ (\mathcal{N}_b, \tilde{\boldsymbol{s}}_{n_b, \tau}) \rightarrow \boldsymbol{x}_{n_b, \tau}$. 
	To perform embedding, for an edge node $n_b \in \mathcal{N}_b$, we build a virtual graph by treating other edge nodes $\mathcal{N}_b \setminus n_b$ as its neighbor nodes.
	Then, as depicted in Fig.~\ref{fig:GNNEncode}, we traverse the edge nodes in $\mathcal{N}_b$ and compute their embedding results one by one. 
	Once an edge node has accomplished embedding, it provides only the embedding results $\boldsymbol{x}_{n_b, \tau}$ for the subsequent embedding processes of the remaining edge nodes.
	For edge node $n_b \in \mathcal{N}_b$, its embedding results $\boldsymbol{x}_{n_b, \tau}$ can be computed by propagating information from its neighbor nodes $\zeta(n_b) = \{ \mathcal{N}_b \setminus n_b \}$ to itself.
	In message passing, edge node $n_b$ aggregates messages from all of its neighbor nodes and computes its embeddings as:
	\begin{equation}
		\label{equ:gnn embedding}
		\boldsymbol{x}_{n_b, \tau}=h_{1}\Big[\sum\nolimits_{n_b^{\prime} \in \zeta(n_b)} f_{1}(\boldsymbol{x}_{n_b^{\prime}, \tau})\Big]+\tilde{\boldsymbol{s}}_{n_b, \tau},
	\end{equation}
	where $h_1(\cdot)$ and $f_1(\cdot)$ are both non-linear transformations implemented by Neural Networks (NNs), combined to express a wide variety of aggregation functions.
	\vthird{During the embedding process, we reuse the same NNs $h_1(\cdot)$ and $f_1(\cdot)$.}

	\item \textbf{\textit{Embedding of eAPs and the edge cluster}}. Similarly, we leverage GNNs to compute an eAP embedding for each eAP $b$, $ \{ \boldsymbol{x}_{n_b, \tau}: n_b \in \mathcal{N}_b \} \rightarrow \boldsymbol{y}_{b, \tau}$, and further an edge cluster embedding for all eAPs, $ \{\boldsymbol{y}_{b, \tau}: b \in \mathcal{B} \} \rightarrow \boldsymbol{z}_{\tau} $. 
	To compute the embedding for eAP $b$ as in (\ref{equ:gnn embedding}), we add an eAP summary node to $\mathcal{N}_b$ and treat all edge nodes in $\mathcal{N}_b$ as its neighbor nodes.
	These eAP summary nodes are also used to store their respective eAP embeddings.
	Then, the eAP embedding for each eAP can be obtained by aggregating messages from all neighboring nodes and computed as (\ref{equ:gnn embedding}).
	In turn, these eAP summary nodes are regarded as the neighbor nodes of an edge cluster summary node, such that (\ref{equ:gnn embedding}) can be used to compute the global embedding as well. 
	Though the embeddings $\boldsymbol{y}_{b, \tau}$ and $\boldsymbol{z}_{\tau}$ are both computed by (\ref{equ:gnn embedding}), different sets of NNs, i.e., (\textit{$\romannumeral1$}) $h_2(\cdot)$, $f_2(\cdot)$ for $\boldsymbol{y}_{b, \tau}$ and (\textit{$\romannumeral2$}) $h_3(\cdot)$, $f_3(\cdot)$ for $\boldsymbol{z}_{\tau}$, are used for non-linear transformations.
\end{itemize} 

\subsubsection{Stepwise Scheduling for Service Orchestration}

The key challenge in encoding service orchestration actions is to deal with the learning and computational complexity of high-dimensional action spaces. 
A direct solution is to maintain a huge policy network and orchestrate all services $\mathcal{W}$ for all edge nodes $\mathcal{N}$ at once based on the embedding results in Sec. \ref{subsubsec:GNN-based System State Encoding}. 
However, in this manner, \textit{KaiS} must choose actions from a large set of combinations $( d_{w, n} \in \mathbb{N}: w \in \mathcal{W}, n \in \mathcal{N})$, \vthird{thus increasing the sample complexity and slowing down the training process}~\cite{silver2016mastering}. 
Besides, too frequent large-scale service orchestration will bring huge system overhead and harm system stability. 

Therefore, we consider stepwise scheduling, which in each frame first selects $H$ high-value edge nodes ($H=2$ in experiments), and then scales services for each of them in a customized action space of a much smaller size $2M+1$. 
Specifically, \textit{KaiS} passes the embedding vectors from Sec. \ref{subsubsec:GNN-based System State Encoding} as inputs to the policy networks, which output a joint orchestration action $\tilde{\boldsymbol{a}}_{\tau} = (\tilde{a}_{\tau}^{\bullet}, \tilde{\boldsymbol{a}}_{\tau}^{\star})$, including (\textit{$\romannumeral1$}) the action of selecting high-value edge nodes $\tilde{a}_{\tau}^{\bullet}$ and (\textit{$\romannumeral2$}) the corresponding joint service scaling action $\tilde{\boldsymbol{a}}_{\tau}^{\star}$.
\begin{figure}[t]%
	\centering
	\includegraphics[width=8.85 cm]{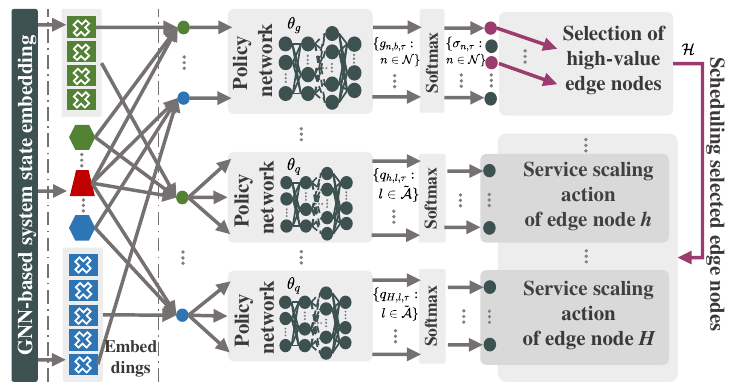}
	\caption{GNN-based learning, i.e., \textit{GPG}, for service orchestration.}
	\label{fig:GPG}
\end{figure}
\begin{itemize}[leftmargin=*]
	\item \textbf{\textit{Selection of high-value edge nodes}}. At each frame, \textit{KaiS} first uses a policy network to select $H (\leq N)$ high-value edge nodes, denoted by action $\tilde{a}_{\tau}^{\bullet}$. As illustrated in Fig.~\ref{fig:GPG}, for edge node $n$ associated with eAP $b$, it computes a value $g_{n, b, \tau} = g(\boldsymbol{x}_{n, \tau}, \boldsymbol{y}_{b, \tau}, \boldsymbol{z}_{\tau})$, where $g(\cdot)$ is a non-linear value-evaluation function implemented by a NN $\theta_g$. The introduction of function $g(\cdot)$ is to map the embedding vectors to a scalar value. The value $g_{n, b, \tau}$ specifies the priority of \textit{KaiS} performing service scaling at edge node $n$. A softmax operation is used to compute the probability $\sigma_{n, \tau}$ of selecting edge node $n$ based on the values $\{ g_{n, b, \tau}: n \in \mathcal{N} \}$:
	\begin{equation}
		\label{equ:softmax}
		\sigma_{n, \tau} = \mathrm{e}^{g_{n, b, \tau}} / \sum\nolimits_{ b^{\prime} \in \mathcal{B} } \sum\nolimits_{n_{b^{\prime}} \in \mathcal{N}_{b^{\prime}}} \mathrm{e}^{g_{n_{b^{\prime}}, b^{\prime}, \tau}}.
	\end{equation}
	According to the probabilities $\{ \sigma_{n, \tau}: n \in \mathcal{N} \}$ for all edge nodes, \textit{KaiS} selects $H$ edge nodes with high probabilities as high-value edge nodes $\mathcal{H}$ to perform service scaling.
	\item \textbf{\textit{Service scaling for high-value edge nodes}}. 
	For a selected high-value edge node $h \in \mathcal{H} \subset \mathcal{N}$, \textit{KaiS} uses an action-evaluation function $q(\cdot)$, implemented by a NN $\theta_q$, to compute a value $q_{h, l, \tau} = q(\boldsymbol{x}_{h, \tau}, \boldsymbol{y}_{b, \tau}, \boldsymbol{z}_{\tau}, l)$ for edge node $h$ performing service scaling $\tilde{a}_{h, \tau} = l$ at frame $\tau$.
	The action space of $l$ is defined as $\tilde{\mathcal{A}} \triangleq (-W, \ldots, w, \ldots, W)$ with size $2W+1$, i.e., $l \in \tilde{\mathcal{A}}$.
	The meaning of $l$ is as follows: (\textit{$\romannumeral1$}) $l = 0$ indicates that all services remain unchanged, (\textit{$\romannumeral2$}) $l = -w$ means deleting a replicate of service $w$, and (\textit{$\romannumeral3$}) $l = w$ specifies adding a replicate of service $w$.
	Particularly, for an invalid service scaling action due to resource limitations of an edge node, \textit{KaiS} always transforms it to $l = 0$. 
	Similarly, we apply a softmax operation on $\{ q_{h, l, \tau}: l \in \tilde{\mathcal{A}} \}$ to compute the probabilities of scaling actions, and choose to perform the action with the highest probability.
	For all high-value edge nodes $\mathcal{H}$, \textit{KaiS} will generate a joint service scaling action $\tilde{\boldsymbol{a}}_{\tau}^{\star} = ( \tilde{a}_{h, \tau}: h \in \mathcal{H} )$ at each frame.
\end{itemize}

While \textit{KaiS} decouples request dispatch and service orchestration, this does not affect our objective of improving $\varPhi^{\prime}$.
\vthird{In fact, by using a regularly updated policy network to provide an appropriate load-balanced edge cluster for orchestration, we implicitly optimize the dispatch policy while optimizing the orchestration.}

To guide \textit{GPG}, \textit{KaiS} generates a reward $\tilde{u}_{\tau} = \mathrm{e}^{ - \sum\nolimits_{n \in \mathcal{N}} \mid \mathcal{Q}_{n, \tau} \mid } $ after each service orchestration at frame $\tau$, where $\mid \mathcal{Q}_{n, \tau} \mid$ is the queue length of unprocessed requests at edge node $n$.
\vthird{By doing so, \textit{GPG} gradually learns to reduce the backlog of unprocessed requests, thereby improving the throughput rate, as we will show in the experiments.}
\textit{KaiS} adopts a policy gradient algorithm for training NNs $\{ f_{i}(\cdot), h_{i}(\cdot) \}_{i=1,2,3}$, $\theta_g$ and $\theta_q$ used in \textit{GPG}.
For clarity, we denote all parameters of these NNs jointly as $\theta^{*}$, all GNN-encoded system states as $\tilde{\boldsymbol{s}}_{\tau}$, the joint service orchestration action as $\tilde{\boldsymbol{a}}_{\tau}$, and the scheduling policy as $\pi_{\theta^{*}}\left( \tilde{\boldsymbol{s}}_{\tau}, \tilde{\boldsymbol{a}}_{\tau} \right)$, i.e., the probability of taking action $\tilde{\boldsymbol{a}}_{\tau}$ when observing state $\tilde{\boldsymbol{s}}_{\tau}$.
At each frame, \textit{KaiS} collects the observation $(\tilde{\boldsymbol{s}}_{\tau}, \tilde{\boldsymbol{a}}_{\tau}, \tilde{u}_{\tau})$ and updates the parameters $\theta^{*}$ using policy gradient:
\begin{equation}
	\label{equ:GNNDRL policy gradient}
	\small
	\theta^{*} \leftarrow \theta^{*} +\alpha \sum_{\tau=1}^{T} \nabla_{\theta^{*}} \log \pi_{\theta^{*}}\left(\tilde{\boldsymbol{s}}_{\tau}, \tilde{\boldsymbol{a}}_{\tau}\right)\left(\sum_{\tau^{\prime}=\tau}^{T} \tilde{u}_{\tau^{\prime}}-\mu_{\tau}\right),
\end{equation}
where $T$ is the length of a \textit{GPG} training episode, $\alpha$ is the learning rate, and $\mu_{\tau}$ is a baseline used to reduce the variance of the policy gradient. 
A method for computing the baseline is setting $\mu_{\tau}$ to the cumulative reward from frame $\tau$ onwards, averaged over all training episodes \cite{greensmith2004variance}.

\section{Implementation Design}
\label{sec:System Design and Implementation}

All services are hosted in the system as \textit{Docker} containers.
In addition, \textit{KaiS} is implemented based on \textit{k8s} v1.18 and \textit{k3s} v1.0 (a lightweight \textit{k8s} for edge) \cite{k3s} in Ubuntu 16.04 using Python 3.6.
\subsection{System Setup}
\label{subsec:End-edge-cloud Computing System Setup}

\begin{itemize}[leftmargin=*]
	\item \textbf{\textit{Requests}}. 
	Real-world workload traces from Alibaba \cite{AlibabaDatasets} are modified and used to generate service requests. 
	We classify the workload requests in that trace into $30$ services.
	Instead of employing real end devices, we implement a \textit{request generator} to generate service requests and then forward them to \textit{k3s master nodes} (eAPs). 
	\item \textbf{\textit{Edge cluster and nodes}}. 
	By default, we set up $5$ \textit{k3s} clusters in different geographic regions of the Google Cloud Platform (GCP) to emulate geographic distribution, each cluster consisting of a \textit{k3s master node} and $8$ \textit{k3s edge nodes}.
	\textit{K3s master nodes} and \textit{k3s edge nodes} use GCP Virtual Machine (VM) configurations ``2 vCPU, 4 GB memory, and 0.3 TB disk'' and ``1-2 vCPU, 2-4 GB memory, and 0.3 TB disk'', respectively. 
	Besides, we use more powerful \textit{k3s master nodes} to accelerate offline training.
	\item \textbf{\textit{Cloud cluster}}. 
	We build a homogeneous 15-VM cluster as the cloud cluster, where each VM is with ``4 vCPU, 16 GB memory, and 1 TB disk''. 
	A \textit{k8s master node} is deployed at one VM to manage others. 
	We handcraft $30$ services with various CPU and memory consumption and store their Docker images in the cloud. 
	\item \textbf{\textit{Network control}}. 
	We intentionally control the network latency of the edge-cloud system, with \textit{Linux TC}, to simulate practical scenarios. Refer to~\cite{hong2019dlion} and~\cite{ barbalace2020edge}, the parameters are set as follows:
	(\textit{$\romannumeral1$}) For the end devices and the edge cluster, the latency is about 20 ms and the bandwidth is about 50 Mbps. 
	(\textit{$\romannumeral2$})  Since the network connection within the edge cluster is through the LAN, the network conditions are better, so the latency is less than 10 ms and the bandwidth is about 1 Gbps. 
	(\textit{$\romannumeral3$}) Since cloud clusters and edge clusters are often geographically distant from each other, WAN is used for the connection. Therefore, the latency is about 100 ms and the bandwidth is about 100 Mbps.
	
\end{itemize}

\subsection{Main Components of KaiS}
\label{subsec:Main Components of KaiS}

\begin{figure}[t]%
	\centering
	\includegraphics[width=8.85 cm]{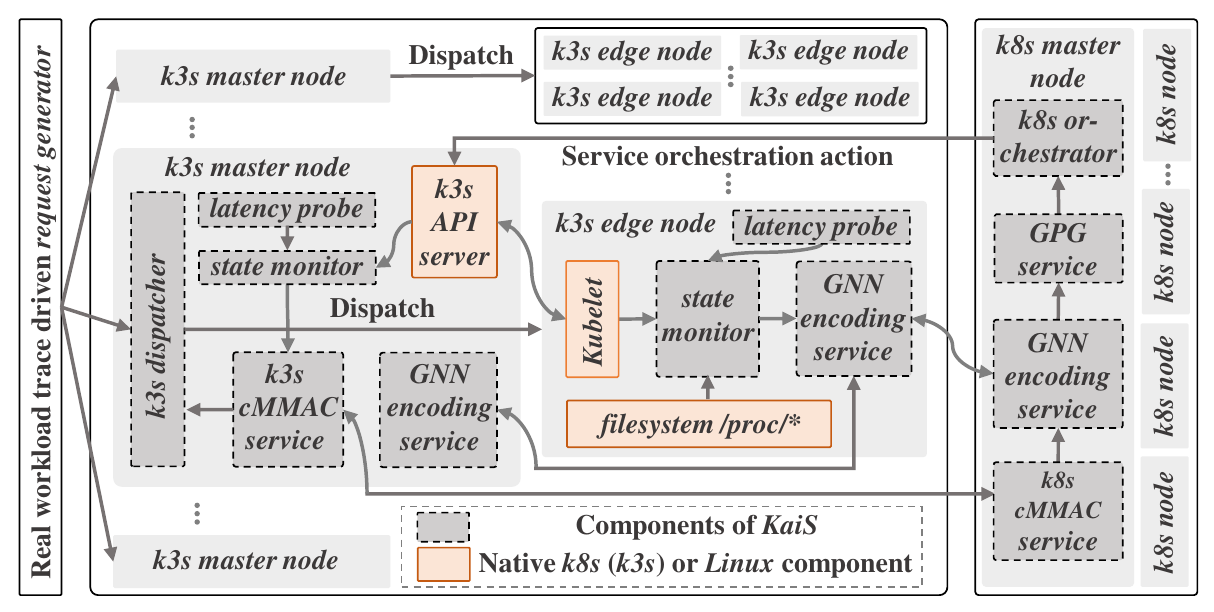}
	\caption{Implementation and prototype design of \textit{KaiS}.}
	\label{fig:ImplementationDesign}
\end{figure}

We decouple \textit{KaiS} into two main parts as shown in Fig.~\ref{fig:ImplementationDesign}. 
\begin{itemize}[leftmargin=*]
	\item \textbf{\textit{Decentralized request dispatchers}}. 
	\textit{KaiS} maintains a \textit{k3s dispatcher} at each \textit{k3s master node} to periodically observe and collect the current system states by a \textit{state monitor} in the following manner. 
	Each \textit{k3s edge node} (\textit{$\romannumeral1$}) runs a \textit{Kubelet} process and (\textit{$\romannumeral2$}) reads the virtual \textit{filesystem /proc/*} in \textit{Linux} to collect the states about \textit{Docker} services and physical nodes.
	Concerning network status, each \textit{k3s edge node} and \textit{k3s master node} host a \textit{latency probe} to measure network latency.
	\textit{State monitors} at \textit{k3s edge nodes} will periodically push the above collected system states to the \textit{state monitor} at the \textit{k3s master node} for fusion.
	To implement \textit{cMMAC}, we deploy a \textit{cMMAC} agent at each \textit{k3s master node} as \textit{k3s cMMAC service} while maintaining a \textit{k8s cMMAC service} at the \textit{k8s master node} to support training. 
	At each scheduling slot $0.25$~s, empirically determined from experiment results as shown in Fig.~\ref{fig:HyperParameters}, the \textit{k3s cMMAC service} at a \textit{k3s master node} computes a dispatch action by observing local states from the \textit{state monitor}, and then notifies the \textit{k3s dispatcher} to execute this dispatch for the current request.
	\item \textbf{\textit{Centralized service orchestrator}}. To implement \textit{GPG}, \textit{KaiS} holds a set of \textit{GNN encoding services} with different GNNs (Sec. \ref{subsubsec:GNN-based System State Encoding}) at \textit{k3s edge nodes}, \textit{k3s master nodes} and \textit{k8s master node}.
	These \textit{GNN encoding services} are communicated with each other and used to compute the embeddings of edge nodes, eAPs (i.e., \textit{k3s master nodes}) and the edge cluster, respectively. 
	Once \textit{KaiS} finishes the GNN-based encoding, the \textit{GNN encoding service} at \textit{k8s master node} will merge all embedding results.
	The remaining parts of \textit{GPG}, i.e., the policy networks, are realized as a \textit{GPG service} and deployed at the \textit{k8s master node}.
	The frame length is set as $100\times$ slots to ensure system stability.
	At each frame, the \textit{GPG service} pulls all embeddings from the \textit{GNN encoding service} and computes the orchestration action.
	Then, the \textit{GPG service} calls the \textit{k8s orchestrator} to communicate with specific \textit{k3s API servers} to accomplish service scaling via \textit{python-k8sclient}.
	Unlike other scaling actions, only when a service is idle, the \textit{k3s API server} can delete it.
	Otherwise, \textit{KaiS} will delay scaling until the condition is met.
\end{itemize}

\subsection{Training Settings}
\label{subsec:Training Settings}

We implement Algorithm \ref{algorithm:global} using TensorFlow 1.14.
The detailed settings are as follows.
\textbf{\textit{cMMAC}}: 
\textit{cMMAC} involves a critic network and an actor policy network. 
Both networks are trained \vthird{using Adam optimizer and a fixed learning rate of $5\times10^{-4}$.}
The critic $\theta_v$ is \vthird{uses a four-layer ReLU NN for parameterization}, where the node sizes of each layer are 256, 128, 64 and 32, respectively. 
The actor $\theta_p$ is implemented using a three-layer ReLU NN, with 256, 128, and 32 hidden units on each layer.
Note that the output layer of the actor uses ReLU+1 as an activation function to ensure that the elements in the original logits are positive. 
\textbf{\textit{GPG}}: 
\textit{GPG} uses (\textit{$\romannumeral1$}) six GNNs, i.e., $\{ f_{i}(\cdot), h_{i}(\cdot) \}_{i=1,2,3}$ and (\textit{$\romannumeral2$}) two policy networks including $\theta_g$ and $\theta_q$. 
Among them, $\{ f_{i}(\cdot), h_{i}(\cdot) \}_{i=1,2,3}$ are implemented with two-hidden-layer NNs with 64 and 32 hidden units on each layer. 
Besides, $\theta_g$ and $\theta_q$ are both three-hidden-layer NNs with node sizes of 128, 64 and 32 from the first layer to the last layer.
All NNs, related to \textit{GPG}, use Adam optimizer with a learning rate of $10^{-3}$ for updates.

For both request dispatch and service orchestration, traditional DRL algorithms cannot cope with continuously arriving service requests.
\vthird{The randomness of arriving requests makes it difficult for the learning algorithm to distinguish between the impact of different service request patterns and the quality of the algorithm's decisions on the change in system performance.}

Besides, the learning-based scheduling policy is bound to make bad decisions in the early stage of training.
Therefore, under the situation that service requests continue to arrive, a scheduling policy that has not been well-trained will inevitably reduce the system throughput, resulting in a backlog of a large number of service requests and untimely processing.
In this case, continuing to spend a lot of time {\color{blue}for exploring} better actions cannot improve the accuracy of the scheduling policy.

To solve the above issues, we adopted a progressive training approach. Inspired by curriculum learning proposed in~\cite{bengio2009curriculum}, we first use simple and short service request sequences for training, and then moderately introduce more sophisticated request sequences step by step, so that the scheduling policy can be gradually improved.

\section{Performance Evaluation}
\label{sec:Performance Evaluation}

Next, we evaluate our design and prototype implementation of \textit{KaiS} in the following aspects:  
(\textit{$\romannumeral1$}) \textbf{\textit{Greedy}} (for dispatch), which schedules each request to the edge node with the lowest resource utilization;
(\textit{$\romannumeral2$}) \textbf{\textit{Native}} (for orchestration), i.e., the default \textit{Horizontal Pod Autoscaler} \cite{k8sNativeScheduler} in \textit{k8s}. {\color{blue}It first periodically observes specific resource metrics in the system and compares them with predefined thresholds, ultimately making decisions based on the gap between the two.}
(\textit{$\romannumeral3$}) \textbf{\textit{GSP-SS}}\cite{Farhadi2019} (for both), assuming that the request arrival rate of each service is known in advance. {\color{blue}It is a two-time-scale framework of joint service placement and request scheduling, and designs a greedy service placement algorithm based on shadow request scheduling computed by a linear program (LP). }
(\textit{$\romannumeral4$}) \textbf{\textit{Firmament}}\cite{Gog2016} (for dispatch), \vthird{designed to assign work (requests) to cloud cluster resources in an optimal manner.} {\color{blue}It is a centralized scheduling method that first summarizes system state observations as graph structured data and then solves the request scheduling problem by a min-cost max-flow (MCMF) based algorithm.}

We consider three main performance metrics:
(\textit{$\romannumeral1$}) \textbf{\textit{Per frame throughput rate }} $\varPhi_f = \left[ \sum\nolimits_{n \in \mathcal{N}} \varUpsilon_{\tau}({\color{blue}\mathcal{Q}^{\prime}_{n}}) + \varUpsilon_{\tau}({\color{blue}\hat{\mathcal{Q}}_{c}}) \right] / $ $\sum\nolimits_{b \in \mathcal{B}} \varUpsilon_{\tau}(\mathcal{Q}_{b})$, which reflects the short-term characteristics of $\varPhi^{\prime}$;
(\textit{$\romannumeral2$}) \textbf{\textit{Scheduling delay}} $\varPhi_d$, the time required for a scheduling action;
(\textit{$\romannumeral3$}) \textbf{\textit{Scheduling cost}} $\varPhi_c$, primarily in terms of network bandwidth consumption, including additional packet forward due to request dispatch, and bandwidth consumption for the edge pulling service \textit{Docker} images from the cloud during service orchestration.
For clarity, we perform the necessary normalization for some metrics, and give their statistical characteristics from the results of multiple experiments.

{\color{blue}
		For request characteristics, they are generated based on the real-world workload traces from Alibaba \cite{AlibabaDatasets}: (\textit{$\romannumeral1$}) The ``start\_time'' in the dataset is used to generate the request arrival time; (\textit{$\romannumeral2$}) The delay requirements of each request is generated by ``end\_time'' minus ``start\_time'' in the dataset; (\textit{$\romannumeral3$}) The ``task\_type'' in the dataset is used to generate the type of each request; (\textit{$\romannumeral4$}) The ``plan\_cpu'' in the dataset is used to generate the CPU requirement of each request; (\textit{$\romannumeral5$}) The ``plan\_mem'' in the dataset is used to generate the memory requirement of each request.
}

\subsection{Learning Ability and Practicability of \textit{KaiS}}

\textit{KaiS} should be able to learn how to cope with request arrivals with underlying statistical patterns and even stochastic request arrivals.

\begin{figure}[t]%
	\centering
	\includegraphics[width=8.85 cm]{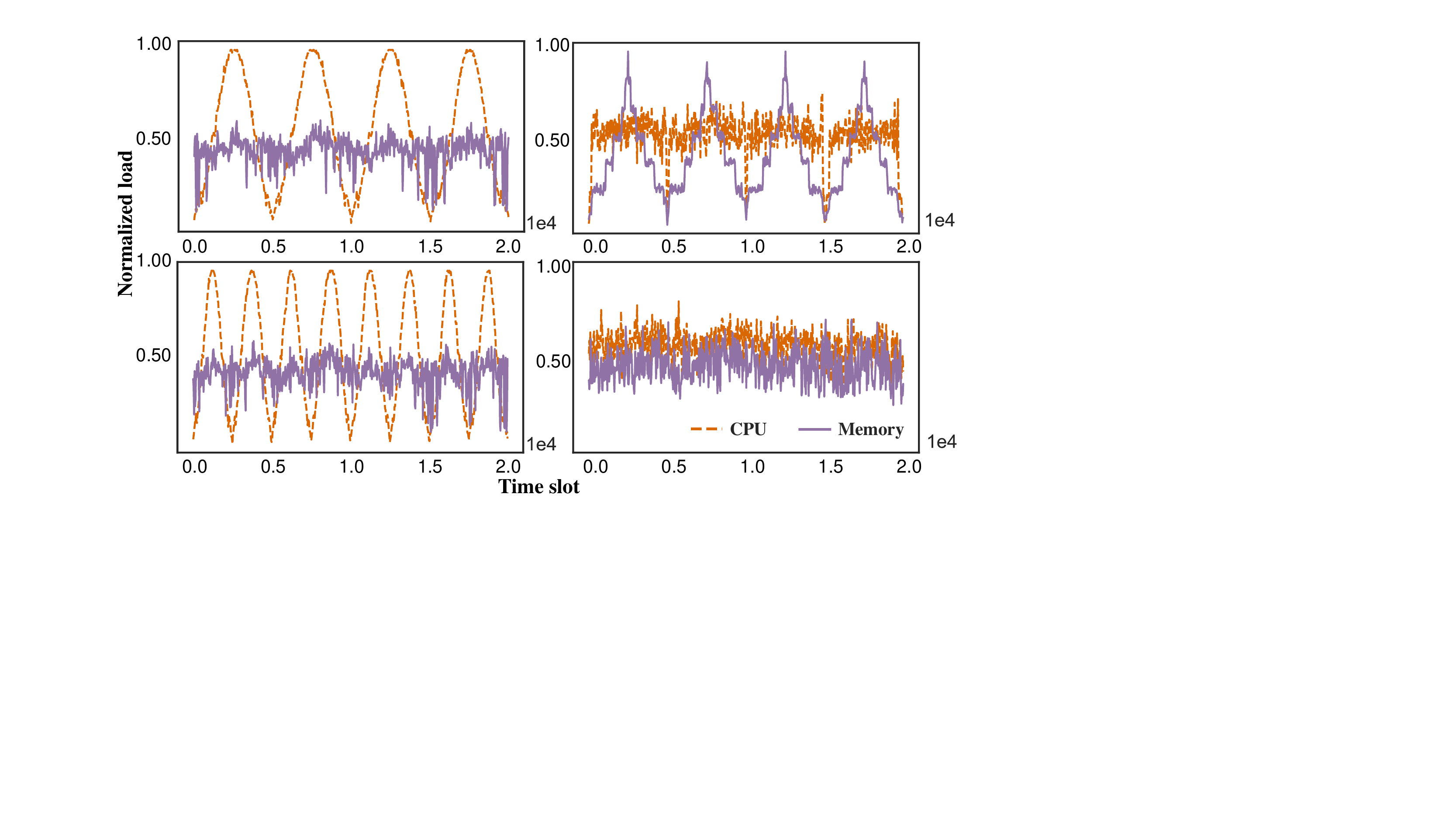}
	\caption{Illustrative examples of 4 request arrival patterns, sampled by the service type: {\color{blue}(a) }$\mathbb{P}_1$ (left top), {\color{blue}(b) }$\mathbb{P}_2$ (right top), {\color{blue}(c) }$\mathbb{P}_3$ (left bottom), {\color{blue}(d) }$\mathbb{P}_4$ (right bottom).}
	\label{fig:RequestPatterns}
\end{figure}

\begin{figure}[t]%
	\centering
	\includegraphics[width=8.85 cm]{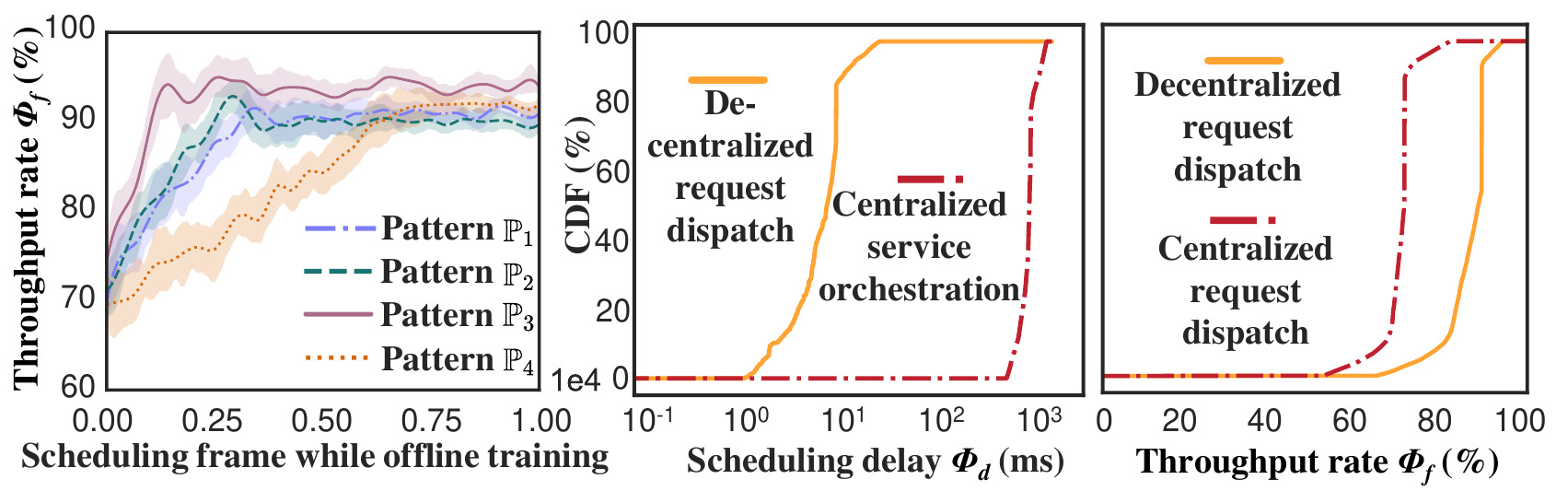}
	\caption{(a) The learning ability of \textit{KaiS} against request arrivals with various patterns (left), and (b) the scheduling delay of \textit{KaiS} performing decentralized request dispatch and centralized service orchestration (central), and (c) the performance of \textit{KaiS} using decentralized and centralized dispatch (right).}
	\label{fig:LearnPatterns}
\end{figure}

\vthird{In Sec. \ref{subsec:End-edge-cloud Computing System Setup}, we sample or clip the workload dataset $\Omega$ to obtain request arrival sequences with four patterns, as shown in Fig.~\ref{fig:RequestPatterns},} viz., (\textit{$\romannumeral1$}) Pattern $\mathbb{P}_1$: periodically fluctuating CPU sum load; (\textit{$\romannumeral2$}) Pattern $\mathbb{P}_2$: periodically fluctuating memory sum load; (\textit{$\romannumeral3$}) Pattern $\mathbb{P}_3$: $\mathbb{P}_1$ with $2\times$ fluctuating frequency; (\textit{$\romannumeral4$}) Pattern $\mathbb{P}_4$: raw stochastic request arrivals clipped from $\Omega$.
{\color{blue} Moreover, CDF (Cumulative Distribution Function) is used in Fig.~\ref{fig:LearnPatterns}(b) and (c), to introduce it we first define the probability of a certain variable $X=x$ as $P(X=x)$ ($x$ is a constant), then the corresponding CDF is $F_X(x)= P(X \leq x)$.}

\begin{figure}[t]%
	\centering
	\includegraphics[width=8.85 cm]{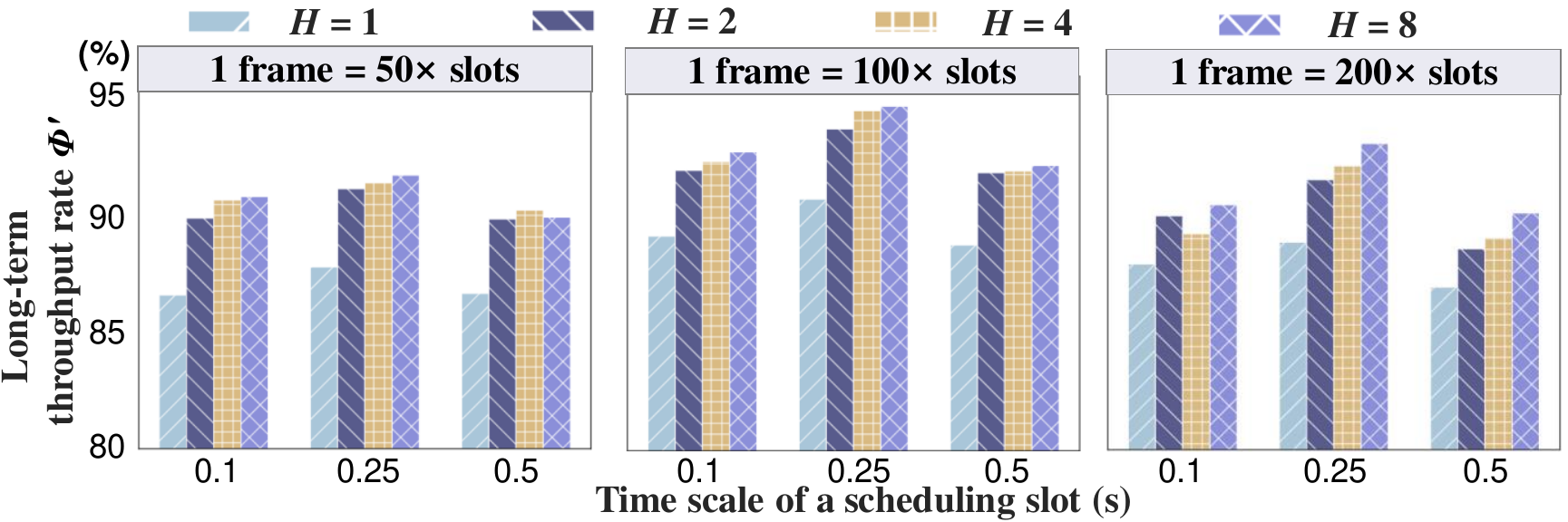}
	\caption{\textit{KaiS}'s scheduling performance under various settings: {\color{blue}(a) 1 frame = 50 $\times$ slots (left),  (b) 1 frame = 100 $\times$ slots (central), (c) 1 frame = 200 $\times$ slots (right).}}
	\label{fig:HyperParameters}
\end{figure}

\begin{itemize}[leftmargin=*]
	\item \textbf{\textit{Learning ability}}.
	Fig.~\ref{fig:LearnPatterns}(a) gives the performance evolution of \textit{KaiS} during training for different request patterns.
	The throughput rate $\varPhi_f$ in all cases is improving over time, \vthird{demonstrating} that \textit{KaiS} can gradually learn to cope with different request patterns.
	In particular, \textit{KaiS} requires experiencing at least $1.2$ times more frames to achieve stable scheduling, when coping with stochastic request arrivals ($\mathbb{P}_4$) rather than others ($\mathbb{P}_{1-3}$).
	Nonetheless, once \textit{KaiS} converges, its scheduling performance gap for requests of different patterns is within $4.5\%$.
	\item \textbf{\textit{Decentralized or centralized dispatch?}}
	Fig.~\ref{fig:LearnPatterns}(b) shows that the scheduling delay of centralized service orchestration is almost $9\times$ \vthird{greater} than that of decentralized request dispatch, while the latter can be completed within around $10$~ms.
	Moreover, we maintain a \textit{cMMAC} agent for each eAP in the cloud to dispatch requests in a centralized manner for comparison.
	From Fig.~\ref{fig:LearnPatterns}(c), we observe that decentralized dispatch can bring higher throughput rates, since centralized dispatch requires additional time to upload local observations ($\hat{\boldsymbol{s}}_{b, t}$) and wait for dispatch decisions.
	However, these extra delays are not trivial for some delay-sensitive service requests.
	\item \textbf{\textit{Two-time-scale scheduling and stepwise orchestration}}. Frequent scheduling may not lead to better performance.
	As shown in Fig.~\ref{fig:HyperParameters}, when a slot is $0.1$s, \textit{cMMAC} agents often experience similar system states in adjacent slots, weakening their learning abilities.
	When a slot is too large ($0.5$s), the untimely dispatch also degrades performance.
	Besides, too frequent service orchestration will result in more scheduling costs and make \textit{cMMAC} agents hard to converge.
	Though selecting more high-value edge nodes for service orchestration at each frame can benefit the throughput, when $H \ge 2$, the improvement is very limited, while a larger $H$ leads to more scheduling {\color{blue}costs}.
	The capability of \textit{KaiS} is affected by the above factors.
	We will show that a default configuration ``$0.25$s (slot), $25$s (frame), $H=2$'' can already yield decent performance compared to baselines. \vthird{It is worth noting that the parameter values can be affected by many factors, such as the nature of the application, request load, edge cluster size, network state, etc. Therefore, they are not generic values and should be adjusted by conducting similar experiments in advance if applied to other scenarios.}

\end{itemize}

\subsection{Impact of Load Balancing}
\begin{figure}[t]%
	\centering
	\includegraphics[width=8.85 cm]{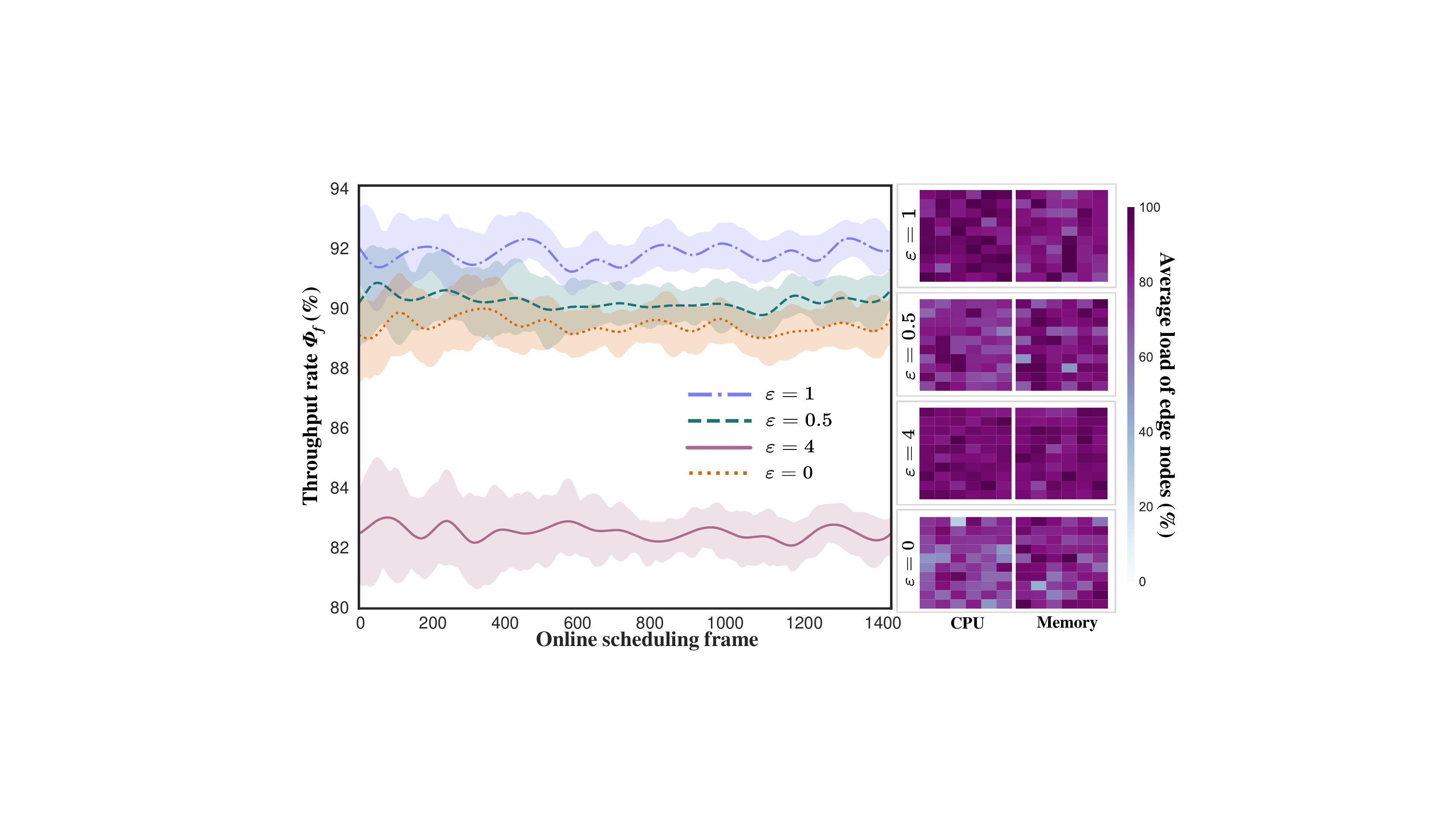}
	\caption{Impact of load balancing on the scheduling performance of \textit{KaiS}: {\color{blue}(a) throughput rate (left) and (b) average load of edge nodes (right).} }
	\label{fig:LoadBalancing}
\end{figure}

In Fig.~\ref{fig:LoadBalancing}, we illustrate the scheduling performance of \textit{KaiS} trained with different settings of $\varepsilon$, that represents the degree of edge load balancing, in $\hat{u}_{b, t} = {\text{e}}^{-\lambda - \varepsilon \nu }$.
\textit{KaiS} achieves the best throughput when $\varepsilon = 1$, while its performance sharply drops when $\varepsilon = 4$.
This performance gap lies in that, when $\varepsilon = 4$, \textit{KaiS} focuses too much on load balancing while in many cases waiving the dispatch options that can tackle requests more efficiently.
Besides, when $\varepsilon = 0$, namely load balancing is not considered, both throughput rate $\varPhi_f$ and load balancing are still better than the case $\varepsilon = 4$.
This fact demonstrates that even if we are not deliberately concerned about load balancing when designing \textit{cMMAC}, \textit{KaiS} can still learn load-balancing policies that are beneficial to improve the throughput. 
Nonetheless, setting a moderate $\varepsilon$ for the reward function can lead \textit{KaiS} to learn more effectively.

\subsection{Effectiveness of Dequeue Strategy}
After that, we design three dequeue strategies for comparison. (\textit{$\romannumeral1$}) \textbf{\textit{First-In-First-Out (FIFO)}}, i.e., all requests are dequeued in the order of the time they enter the request queue. (\textit{$\romannumeral2$}) \textbf{\textit{Latency-Greedy}}, i.e., since the requests are latency-sensitive, the requests in the request queue are sorted according to the delay requirements, and the closer the timeout is, the higher the queueing priority is. (\textit{$\romannumeral3$}) \textbf{\textit{Discounted-Experience}}, i.e., by calculating the predicted completion time of various types of requests based on discounted experience, the strategy prioritizes the {\color{blue}dequeuing} of the request whose predicted completion time is close to the deadline.

\begin{figure}[t]%
	\centering
	\includegraphics[width=8.85 cm]{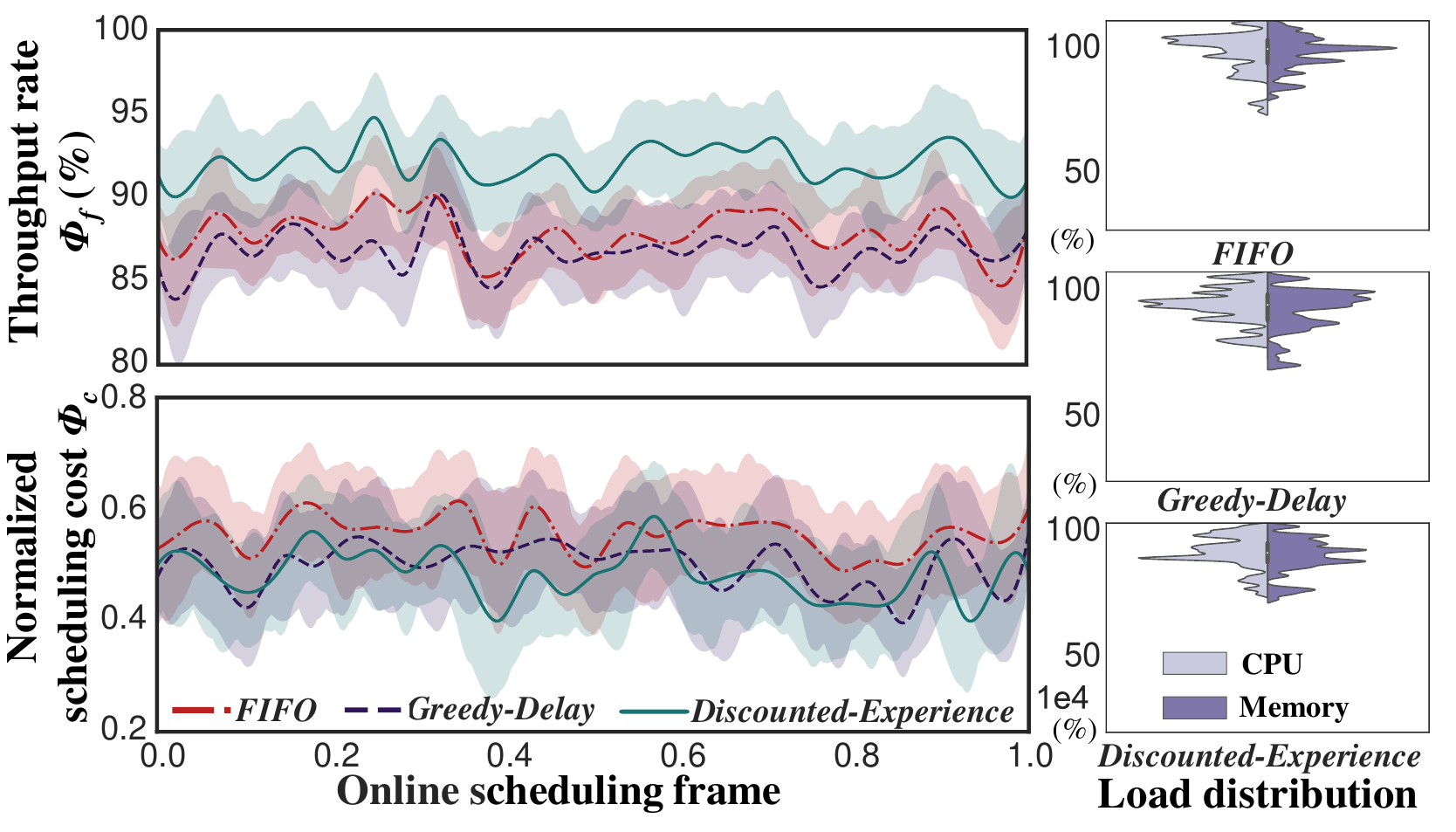}
	\caption{The performance of different dequeue strategies in terms of {\color{blue}(a)} throughput rate (top left) and {\color{blue}(b)} cost (bottom left), and {\color{blue}(c)} the percentage of nodes with different resource occupancy rates (right).}
	\label{fig:Dequeue_Strategy}
\end{figure}

As shown in Fig.~\ref{fig:Dequeue_Strategy}, we compare the effectiveness of the three dequeue strategies mentioned above. It can be found that the difference between the three strategies in terms of cost is relatively small and roughly at the same level. However, in terms of throughput, the performance of \textit{FIFO} is close to \textit{Greedy}-\textit{Delay} and the \textit{Discounted}-\textit{Experience} has an obvious advantage. 
The reasons for the above experimental results are as follows.
For the strategy of \textit{FIFO}, the arrival time of the request cannot correctly reflect the processing priority of the request \vthird{because different requests have varying delay requirements. Therefore, \textit{FIFO} cannot effectively handle requests with extreme latency requirements, resulting in resource waste and decreased throughput.}
In addition, \textit{Greedy}-\textit{Delay} also has some shortcomings. For the requests that are about to exceed the delay requirement, they may not be completed on time even if they are immediately dequeued due to the short remaining time. However, these requests have high priorities under the dequeue strategy of \textit{Greedy}-\textit{Delay}, and processing them only results in wasted resources and has a negative effect on the throughput. 
In contrast, the dequeue strategy of \textit{Discounted-Experience} we designed has obvious advantages. It can avoid the disadvantages mentioned in the other two dequeue strategies and reduce resource waste by reasonably arranging the queueing time of each request.

\subsection{Role of GNN-based Service Orchestration}

Next, \vthird{we first combine request arrival sequences of four patterns to construct a long series, to evaluate \textit{GPG}'s ability to respond to request arrivals with fluctuating patterns.}
Note that these long request arrival sequences are constructed to reflect scenarios with high variability.
\begin{figure}[t]%
	\centering
	\includegraphics[width=8.85 cm]{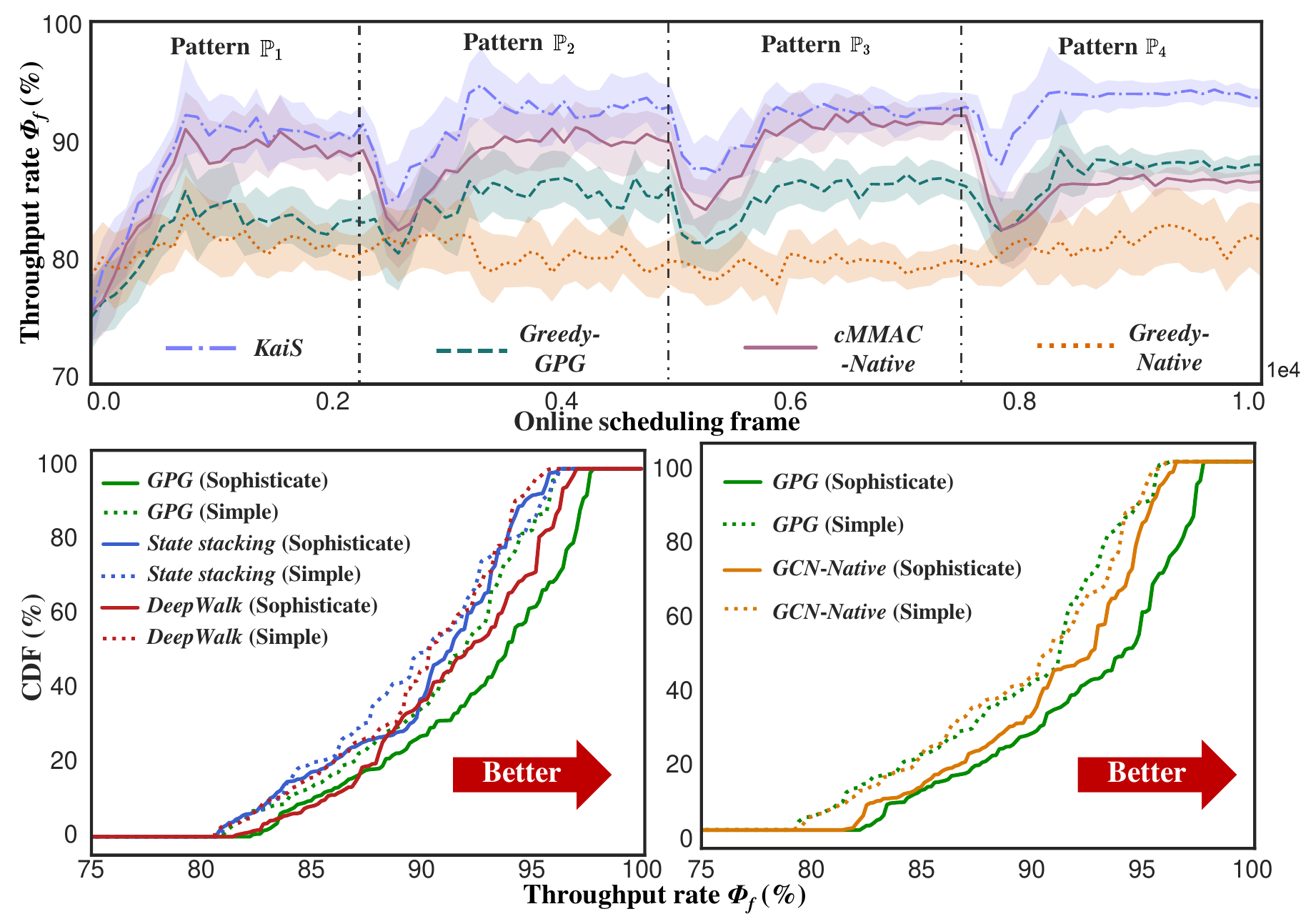}
	\caption{(a) The ability of \textit{GPG} to respond to pattern-fluctuating request arrivals (top). (b, c) In different system scales, compared with other methods of state encoding (left) and with or without custom modifications (right).}
	\label{fig:GNNRole}
\end{figure}
\begin{itemize}[leftmargin=*]
	\item \textbf{\textit{Coping with stochastic request arrivals}}. In Fig.~\ref{fig:GNNRole}(a), presented the scheduling performance of \textit{KaiS}, \textit{Greedy}-\textit{GPG}, \textit{cMMAC}-\textit{Native} and \textit{Greedy}-\textit{Native} to the scenarios with high variability.
	The results \vthird{demonstrate} the following points: (\textit{$\romannumeral1$}) \textit{KaiS} achieves higher average throughput rate than the closest competing baselines, and particularly, whenever the request arrival pattern changes, \textit{KaiS} can still quickly learn a policy adapted to the new pattern; (\textit{$\romannumeral2$}) For patterns $\mathbb{P}_{1,2,3}$, \textit{cMMAC}-\textit{Native} can achieve scheduling performance close to \textit{KaiS}, the reason of which lies in that an efficient request dispatch algorithm, e.g., \textit{cMMAC}, can already address the request arrivals with obvious patterns; (\textit{$\romannumeral3$}) For the sophisticated pattern $\mathbb{P}_4$, i.e., requests are arriving stochastically as the raw traces $\Omega$, due to the lack of service orchestration to adaptively release and capture the global system resources, the performance of \textit{cMMAC}-\textit{Native} and \textit{Greedy}-\textit{Native} deteriorates.

	\item \textbf{\textit{GNN-based encoding against other methods}}. 
	We show in Fig.~\ref{fig:GNNRole}(b) the role of \textit{GNN-based system state encoding} (Sec. \ref{subsubsec:GNN-based System State Encoding}).
	For evaluation, we build two edge clusters with different system scales: (\textit{$\romannumeral1$}) a default setting introduced in Sec. \ref{subsec:End-edge-cloud Computing System Setup}; (\textit{$\romannumeral2$}) a complex setting with $10$ \textit{k3s master nodes}, each of which manages $3$-$15$ heterogeneous \textit{k3s edge nodes} ($100$ in total).	
	In addition, we select two state encoding methods, \textit{State stacking} and \textit{DeepWalk}, for comparison. In detail, \textit{State stacking} is directly stacking system states into flat vectors. 
	\textit{DeepWalk} differs from \textit{GPG}'s nonlinear mapping based on neural networks, and it uses a random walk strategy to generate a low-dimensional representation of the graph~\cite{perozzi2014deepwalk}, which has been applied in research work about the edge-cloud network~\cite{kong2020mobile}.
	From Fig.~\ref{fig:GNNRole}(b), we can observe the following points:
	(\textit{$\romannumeral1$}) Under different system scales, \textit{GPG} is superior to other methods, with a small gap for \textit{DeepWalk} but a large gap for \textit{State stacking};
	(\textit{$\romannumeral2$}) As the system scale becomes larger, the gap between different state encoding methods becomes larger;
	(\textit{$\romannumeral3$}) \textit{State stacking} does not show a significant improvement at larger system scales, which is because the larger scale will bring more complex states, but it cannot accurately extract state information resulting in poorly trained models, which significantly reduces the accuracy of inference and overall system efficiency.
	In summary, GNN-based encoding can significantly reduce \textit{KaiS} dependence on the model complexity of NNs, which is key to efficient and fast learning. 
	Further, it embeds the network latency and the system structure information, assisting \textit{KaiS} scale to large-scale edge-cloud network.
	
	\item \textbf{\textit{Significance of custom modifications}}. 	
	\textit{GPG} described in Sec. \ref{subsec:GNN-based DRL for Service Orchestration} is based on the generic Graph Convolutional Network (GCN) with custom modifications for the edge-cloud network, and we designed \textit{GCN}-\textit{Native} for comparison to check whether the adopted custom modifications can improve performance.
	In detail, \textit{GCN}-\textit{Native} encodes the graph data composed of global native system states directly based on the generic GCN, i.e., different from the message aggregation and multi-level embedding of \textit{GPG}.
	As shown in Fig.~\ref{fig:GNNRole}(c), there is no significant difference between \textit{GPG} and \textit{GCN}-\textit{Native} under the small-scale edge-cloud system, but \textit{GPG} \vthird{outperforms} with the expansion of the system scale.
	The above experimental results demonstrate that the custom modifications can help \textit{KaiS} efficiently extract information from environments, which can enhance the adaptability of \textit{KaiS} to large-scale edge-cloud networks. 
	
\end{itemize}

\subsection{Performance Comparison with Baselines}

\begin{figure}[t]%
	\centering
	\includegraphics[width=8.85 cm]{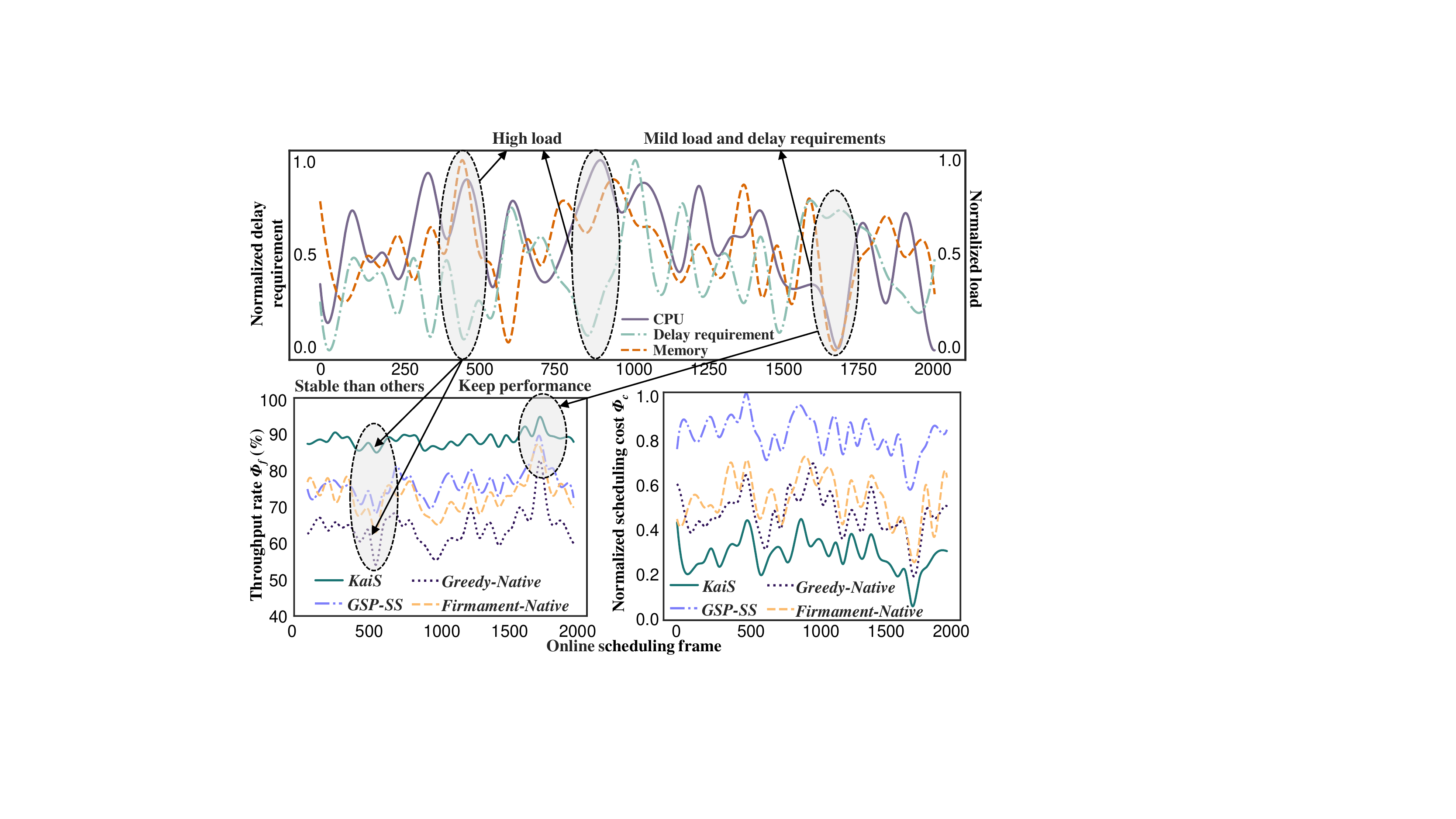}
	\caption{(a) Under stochastic request arrivals (top), (b, c) the performance of \textit{KaiS} against baselines in terms of throughput rate (left) and cost (right).}
	\label{fig:Baselines}
\end{figure}

To evaluate \textit{KaiS}, we need to consider both scheduling performance and cost. 
We clip the workload dataset $ \Omega $ to obtain $50$ request arrival sequences with the same length and use them to evaluate \textit{KaiS}. 
From Fig.~\ref{fig:Baselines}(b-c), we observe that in almost all cases, regardless of how the loads and the delay requirements of requests fluctuate (Fig.~\ref{fig:Baselines}(a)), \textit{KaiS} yields a $14.3 \%$ higher throughput rate $\varPhi_f$ and a $34.7 \%$ lower scheduling cost $\varPhi_c$ than the closest competing baselines.

Particularly, when the request loads and delay requirements are mild at some frames, the scheduling performance $\varPhi_f$ of \textit{GSP-SS} can be very close to that of \textit{KaiS}. 
However, in contrast to \textit{KaiS}, the scheduling performance of \textit{GSP-SS} degrades during frames with high loads: as it does not understand the system capability to process requests, when the request load level is high, it cannot load balancing the edge cluster to apportion these loads, thereby narrowing available scheduling spaces.
Besides, (\textit{$\romannumeral1$}) \textit{KaiS} adopts two-time-scale scheduling, and unlike \textit{GSP-SS} performing large-scale orchestration at each frame, (\textit{$\romannumeral2$}) it only selects a fixed number $H$ of high-value edge nodes to perform service orchestration limited by $\tilde{\mathcal{A}}$.
Hence, the scheduling cost of \textit{KaiS} is bounded in each frame, thereby reducing the overall cost, as shown in Fig.~\ref{fig:Baselines}(b).

\section{Related Work}
\label{sec:Related Work}

Though existing optimization studies explore the upper bounds of scheduling performance, they are not applicable to practical deployment environments (e.g., \textit{k8s}) due to various model assumptions.
Besides, there exists no system design studies to accommodate decentralized request dispatch.

\subsection{Theoretical Analysis}
Most of the current theoretical studies explore one of the two problems, i.e., request dispatch and service orchestration. 
For request dispatch, \cite{huang2020task, jovsilo2020computation, dong2019computation} provide scheduling algorithms for different optimization objectives in edge computing scenarios, but some key system parameters such as task size, computing capacity of the device, and network communication capability in their simulation models are set with static constants or ranges of values.
For service orchestration, \cite{liang2019interaction,kayal2019distributed, badri2019energy } \vthird{combine different algorithmic design perspectives} to provide solutions, but both of them cannot take full advantage of centralized cloud clustering and cannot sufficiently consider hardware resource constraints.

Many studies, e.g.,~\cite{Zou2020, Chen2016f}, have given scheduling solutions for offloading stochastic computation or service requests, which complement our study. 
The studies of~\cite{Farhadi2019, Poularakis2019, Ma2020} set a theoretical basis {\color{blue}for} jointly optimizing request dispatch and service orchestration.
However, the proposed one-shot scheduling optimization in~\cite{Poularakis2019, Ma2020} cannot address continuously arriving service requests, i.e., without considering the long-term impact of scheduling.
In~\cite{Farhadi2019}, the authors propose to perform optimization on two different time scales to maximize the number of requests completed in each schedule.
Nevertheless, the long-term optimization in~\cite{Farhadi2019} relies on the accurate prediction of future service requests, which is difficult to achieve in practice.
\vthird{Last but not least, these studies~\cite{Farhadi2019, Poularakis2019, Ma2020} are not practically applicable} since: 
(\textit{$\romannumeral1$}) They assume that the computing resources, network requirements, or the processing time for specific requests can be accurately modeled or predicted;
(\textit{$\romannumeral2$}) The dispatch is scheduled in a centralized manner, while it must take extra time to wait for the aggregation of context information across the entire system.

\subsection{System Design}

Many efficient schedulers have been developed for \textit{k8s}-based cloud clusters.
These studies either schedule all tasks through minimum cost maximum flow optimization for general workloads~\cite{Gog2016} or exploit domain-specific knowledge of, e.g., deep learning, to improve overall cluster utilization for specific workloads~\cite{Xiao2018b}.
However, they cannot accommodate decentralized request dispatch at the edge, since their schedulers are deployed {\color{blue}in} the cloud in a centralized fashion.
The scheduler proposed in~\cite{Haja2019} orchestrates services by periodically measuring the latency between edge nodes to estimate whether the expected processing delay of service requests can meet requirements.
The study most related to ours is~\cite{Rossi2020}, which uses model-based RL to deal with the service orchestration and is compatible with geographically distributed edge clusters.
Nonetheless, neither~\cite{Haja2019} nor~\cite{Rossi2020} consider dispatching of requests at the edge clusters.
{\color{blue}As far as we know, the current studies about system design for \textit{k8s} only focus on }
the issue of service orchestration, while ignoring the issue of request dispatch. Therefore, there is a lack of system design related to the collaborative optimization of requests and services.

\section{Conclusion}
\label{sec:Conclusion}

Leveraging \textit{k8s} to seamlessly merge the distributed edge and the cloud is the future of the edge-cloud network.
In this paper, we have introduced \textit{KaiS}, a scheduling framework integrated with tailored learning algorithms for \textit{k8s}-based edge-cloud network, that dynamically learns scheduling policies for request dispatch and service orchestration to improve the long-term system throughput rate.
To this end, we tailor learning algorithms for \textit{KaiS}, including a coordinated multi-agent actor-critic algorithm designed for decentralized request dispatch combined with a dequeue strategy based on discounted experience, and a GNN-based policy gradient algorithm for centralized service orchestration. In order to verify the effectiveness of \textit{KaiS}, we conduct systematic experiments in the edge-cloud network.
Our results show the behavior of \textit{KaiS} across different scenarios and demonstrate that \textit{KaiS} can at least enhance the average system throughput rate by $15.9\%$ while reducing scheduling {\color{blue}costs} by $38.4\%$.
In addition, by modifying the scheduling action spaces and reward functions, \textit{KaiS} is also applicable to other scheduling optimization goals, such as minimizing the long-term system overhead.

\bibliographystyle{IEEEtran}
\bibliography{TON2021} 

\begin{thebibliography}{10}
\providecommand{\url}[1]{#1}
\csname url@samestyle\endcsname
\providecommand{\newblock}{\relax}
\providecommand{\bibinfo}[2]{#2}
\providecommand{\BIBentrySTDinterwordspacing}{\spaceskip=0pt\relax}
\providecommand{\BIBentryALTinterwordstretchfactor}{4}
\providecommand{\BIBentryALTinterwordspacing}{\spaceskip=\fontdimen2\font plus
\BIBentryALTinterwordstretchfactor\fontdimen3\font minus
  \fontdimen4\font\relax}
\providecommand{\BIBforeignlanguage}[2]{{%
\expandafter\ifx\csname l@#1\endcsname\relax
\typeout{** WARNING: IEEEtran.bst: No hyphenation pattern has been}%
\typeout{** loaded for the language `#1'. Using the pattern for}%
\typeout{** the default language instead.}%
\else
\language=\csname l@#1\endcsname
\fi
#2}}
\providecommand{\BIBdecl}{\relax}
\BIBdecl

\bibitem{kais}
Y.~Han, S.~Shen, X.~Wang, S.~Wang, and V.~Leung, ``{Tailored Learning-Based
  Scheduling for Kubernetes-Oriented Edge-Cloud System},'' in \emph{IEEE
  INFOCOM}, 2021.

\bibitem{Shi2016}
W.~Shi, J.~Cao \emph{et~al.}, ``{Edge Computing: Vision and Challenges},''
  \emph{IEEE Internet Things J.}, vol.~3, no.~5, pp. 637--646, Oct. 2016.

\bibitem{Burns2016}
B.~Burns, B.~Grant, D.~Oppenheimer, E.~Brewer, and J.~Wilkes, ``{Borg, Omega,
  and Kubernetes},'' \emph{Commun. ACM}, Apr. 2016.

\bibitem{kubeedge}
\BIBentryALTinterwordspacing
``{KubeEdge}: Kubernetes native edge computing framework (project under
  {CNCF}).'' [Online]. Available: \url{https://github.com/kubeedge/kubeedge}
\BIBentrySTDinterwordspacing

\bibitem{openyurt}
\BIBentryALTinterwordspacing
``{OpenYurt}: Extending your native kubernetes to edge.'' [Online]. Available:
  \url{https://github.com/alibaba/openyurt}
\BIBentrySTDinterwordspacing

\bibitem{baetyl}
\BIBentryALTinterwordspacing
``Baetyl: Extend cloud computing, data and service seamlessly to edge
  devices.'' [Online]. Available: \url{https://github.com/baetyl/baetyl}
\BIBentrySTDinterwordspacing

\bibitem{Wang2020}
X.~Wang, Y.~Han, V.~C. Leung, D.~Niyato, X.~Yan, and X.~Chen, ``{Convergence of
  Edge Computing and Deep Learning: A Comprehensive Survey},'' \emph{IEEE
  Commun. Surv. Tutor.}, vol.~22, no.~2, pp. 869--904, 2020.

\bibitem{Tan2017}
H.~Tan, Z.~Han, X.-y. Li, and F.~C. Lau, ``{Online job dispatching and
  scheduling in edge-clouds},'' in \emph{IEEE INFOCOM}, 2017.

\bibitem{Pasteris2019}
S.~Pasteris, S.~Wang, M.~Herbster, and T.~He, ``{Service Placement with
  Provable Guarantees in Heterogeneous Edge Computing Systems},'' in \emph{IEEE
  INFOCOM}, 2019.

\bibitem{Farhadi2019}
V.~Farhadi, F.~Mehmeti, T.~He, T.~L. Porta, H.~Khamfroush, S.~Wang, and K.~S.
  Chan, ``{Service Placement and Request Scheduling for Data-intensive
  Applications in Edge Clouds},'' in \emph{IEEE INFOCOM}, 2019.

\bibitem{Poularakis2019}
K.~Poularakis, J.~Llorca, A.~M. Tulino, I.~Taylor, and L.~Tassiulas, ``{Joint
  Service Placement and Request Routing in Multi-cell Mobile Edge Computing
  Networks},'' in \emph{IEEE INFOCOM}, 2019.

\bibitem{Ma2020}
X.~Ma, S.~Wang \emph{et~al.}, ``{Cooperative Service Caching and Workload
  Scheduling in Mobile Edge Computing},'' in \emph{IEEE INFOCOM}, 2020.

\bibitem{Ayala-Romero2019}
J.~A. Ayala-Romero, A.~Garcia-Saavedra, M.~Gramaglia, X.~Costa-Perez,
  A.~Banchs, and J.~J. Alcaraz, ``{vrAIn: A Deep Learning Approach Tailoring
  Computing and Radio Resources in Virtualized RANs},'' in \emph{ACM MobiCom},
  2019.

\bibitem{sutton2018reinforcement}
R.~S. Sutton \emph{et~al.}, \emph{Reinforcement learning: An introduction}.

\bibitem{mao2019learning}
H.~Mao, M.~Schwarzkopf, S.~B. Venkatakrishnan, Z.~Meng, and M.~Alizadeh,
  ``Learning scheduling algorithms for data processing clusters,'' in \emph{ACM
  SIGCOMM}, 2019, pp. 270--288.

\bibitem{Ren2019b}
J.~Ren, D.~Zhang, S.~He, Y.~Zhang, and T.~Li, ``{A Survey on End-Edge-Cloud
  Orchestrated Network Computing Paradigms},'' \emph{ACM Comput. Surv.},
  vol.~52, no.~6, pp. 1--36, Oct. 2019.

\bibitem{Mnih2015}
V.~Mnih \emph{et~al.}, ``{Human-level control through deep reinforcement
  learning},'' \emph{Nature}, vol. 518, no. 7540, pp. 529--533, Feb. 2015.

\bibitem{lillicrap2015continuous}
T.~P. Lillicrap \emph{et~al.}, ``{Continuous control with deep reinforcement
  learning},'' \emph{arXiv preprint arXiv:1509.02971}, 2015.

\bibitem{WangMADRL}
F.~Wang, F.~Wang, J.~Liu, R.~Shea, and L.~Sun, ``{Intelligent Video Caching at
  Network Edge : A Multi-Agent Deep Reinforcement Learning Approach},'' in
  \emph{IEEE INFOCOM}, 2020.

\bibitem{MARLSurvey}
L.~{Busoniu}, R.~{Babuska}, and B.~{De Schutter}, ``A comprehensive survey of
  multiagent reinforcement learning,'' \emph{IEEE Trans. Syst., Man, Cybern. C,
  Appl. Rev.}, vol.~38, no.~2, pp. 156--172, 2008.

\bibitem{Zhang2020}
Z.~Zhang, P.~Cui, and W.~Zhu, ``{Deep Learning on Graphs: A Survey},''
  \emph{IEEE Trans. Knowl. Data Eng. (Early Access)}, 2020.

\bibitem{Silver2014}
D.~Silver, G.~Lever, N.~Heess, T.~Degris, D.~Wierstra, and M.~Riedmiller,
  ``{Deterministic policy gradient algorithms},'' in \emph{ICML}, 2014.

\bibitem{chen2021multitask}
J.~Chen, Y.~Yang, C.~Wang, H.~Zhang, C.~Qiu, and X.~Wang, ``Multitask
  offloading strategy optimization based on directed acyclic graphs for edge
  computing,'' \emph{IEEE Internet of Things Journal}, vol.~9, no.~12, pp.
  9367--9378, 2021.

\bibitem{liu2022hastening}
Z.~Liu, J.~Song, C.~Qiu, X.~Wang, X.~Chen, Q.~He, and H.~Sheng, ``Hastening
  stream offloading of inference via multi-exit dnns in mobile edge
  computing,'' \emph{IEEE Transactions on Mobile Computing}, 2022.

\bibitem{lv2022microservice}
W.~Lv, Q.~Wang, P.~Yang, Y.~Ding, B.~Yi, Z.~Wang, and C.~Lin, ``Microservice
  deployment in edge computing based on deep q learning,'' \emph{IEEE Trans
  Parallel Distrib Syst}, 2022.

\bibitem{ding2022kubernetes}
Z.~Ding, S.~Wang, and C.~Jiang, ``Kubernetes-oriented microservice placement
  with dynamic resource allocation,'' \emph{IEEE Trans. on Cloud Comput.},
  no.~01, pp. 1--1, 2022.

\bibitem{wang2021multi}
X.~Wang, Z.~Ning, and S.~Guo, ``Multi-agent imitation learning for pervasive
  edge computing: A decentralized computation offloading algorithm,''
  \emph{IEEE Trans Parallel Distrib Syst}, vol.~32, no.~2, pp. 411--425, 2021.

\bibitem{jovsilo2020computation}
S.~Jo{\v{s}}ilo and G.~D{\'a}n, ``Computation offloading scheduling for
  periodic tasks in mobile edge computing,'' \emph{IEEE/ACM Trans. Netw.},
  vol.~28, no.~2, pp. 667--680, 2020.

\bibitem{AlibabaDatasets}
\BIBentryALTinterwordspacing
``Aliababa-clusterdata.'' [Online]. Available:
  \url{https://github.com/alibaba/clusterdata}
\BIBentrySTDinterwordspacing

\bibitem{Chen2016f}
X.~Chen, L.~Jiao, W.~Li, and X.~Fu, ``{Efficient Multi-User Computation
  Offloading for Mobile-Edge Cloud Computing},'' \emph{IEEE/ACM Trans. Netw.},
  vol.~24, no.~5, pp. 2795--2808, 2016.

\bibitem{mnih2015human}
V.~Mnih, K.~Kavukcuoglu, D.~Silver, A.~A. Rusu, J.~Veness, M.~G. Bellemare,
  A.~Graves, M.~Riedmiller, A.~K. Fidjeland, G.~Ostrovski \emph{et~al.},
  ``Human-level control through deep reinforcement learning,'' \emph{nature},
  vol. 518, no. 7540, pp. 529--533, 2015.

\bibitem{silver2016mastering}
D.~Silver \emph{et~al.}, ``Mastering the game of go with deep neural networks
  and tree search,'' \emph{Nature}, vol. 529, no. 7587, pp. 484--489, 2016.

\bibitem{greensmith2004variance}
E.~Greensmith \emph{et~al.}, ``Variance reduction techniques for gradient
  estimates in reinforcement learning,'' \emph{J. Mach. Learn. Res.}, 2004.

\bibitem{k3s}
\BIBentryALTinterwordspacing
``Lightweight kubernetes.'' [Online]. Available:
  \url{https://github.com/rancher/k3s}
\BIBentrySTDinterwordspacing

\bibitem{hong2019dlion}
R.~Hong and A.~Chandra, ``Dlion: Decentralized distributed deep learning in
  micro-clouds,'' in \emph{USENIX HotCloud}, 2019.

\bibitem{barbalace2020edge}
A.~Barbalace, M.~L. Karaoui, W.~Wang, T.~Xing, P.~Olivier, and B.~Ravindran,
  ``Edge computing: the case for heterogeneous-isa container migration,'' in
  \emph{ACM VEE}, 2020, pp. 73--87.

\bibitem{bengio2009curriculum}
Y.~Bengio, J.~Louradour, R.~Collobert, and J.~Weston, ``Curriculum learning,''
  in \emph{ACM ICML}, 2009, pp. 41--48.

\bibitem{k8sNativeScheduler}
\BIBentryALTinterwordspacing
``{K8s documentation: horizontal pod autoscaler}.'' [Online]. Available:
  \url{https://kubernetes.io/docs/tasks/run-application/horizontal-pod-autoscale/}
\BIBentrySTDinterwordspacing

\bibitem{Gog2016}
I.~Gog, M.~Schwarzkop \emph{et~al.}, ``{Firmament: Fast, centralized cluster
  scheduling at scale},'' in \emph{USENIX OSDI}, 2016.

\bibitem{perozzi2014deepwalk}
B.~Perozzi, R.~Al-Rfou, and S.~Skiena, ``Deepwalk: Online learning of social
  representations,'' in \emph{ACM SIGKDD}, 2014, pp. 701--710.

\bibitem{kong2020mobile}
X.~Kong, S.~Tong, H.~Gao, G.~Shen, K.~Wang, M.~Collotta, I.~You, and S.~Das,
  ``Mobile edge cooperation optimization for wearable internet of things: a
  network representation-based framework,'' \emph{IEEE Trans. Ind. Informat.},
  2020.

\bibitem{huang2020task}
X.~Huang, R.~Yu, S.~Xie, and Y.~Zhang, ``Task-container matching game for
  computation offloading in vehicular edge computing and networks,'' \emph{IEEE
  Trans. Intell. Transp. Syst.}, 2020.

\bibitem{dong2019computation}
L.~Dong, M.~N. Satpute, J.~Shan, B.~Liu, Y.~Yu, and T.~Yan, ``Computation
  offloading for mobile-edge computing with multi-user,'' in \emph{IEEE
  ICDCS}.\hskip 1em plus 0.5em minus 0.4em\relax IEEE, 2019, pp. 841--850.

\bibitem{liang2019interaction}
Y.~Liang, J.~Ge, S.~Zhang, J.~Wu, L.~Pan, T.~Zhang, and B.~Luo,
  ``Interaction-oriented service entity placement in edge computing,''
  \emph{IEEE Trans. Mob. Comput. (Early Access)}, 2019.

\bibitem{kayal2019distributed}
P.~Kayal and J.~Liebeherr, ``Distributed service placement in fog computing: An
  iterative combinatorial auction approach,'' in \emph{IEEE ICDCS}.\hskip 1em
  plus 0.5em minus 0.4em\relax IEEE, 2019, pp. 2145--2156.

\bibitem{badri2019energy}
H.~Badri, T.~Bahreini, D.~Grosu, and K.~Yang, ``Energy-aware application
  placement in mobile edge computing: A stochastic optimization approach,''
  \emph{IEEE Trans. Parallel Distrib. Syst.}, vol.~31, no.~4, pp. 909--922,
  2019.

\bibitem{Zou2020}
J.~Zou, T.~Hao, C.~Yu, and H.~Jin, ``{A3C-DO: A Regional Resource Scheduling
  Framework based on Deep Reinforcement Learning in Edge Scenario},''
  \emph{IEEE Trans. Comput.}, vol.~70, no.~2, pp. 228--239, Feb. 2021.

\bibitem{Xiao2018b}
W.~Xiao, Z.~Han \emph{et~al.}, ``{Gandiva: Introspective cluster scheduling for
  deep learning},'' in \emph{USENIX OSDI}, 2018.

\bibitem{Haja2019}
D.~Haja, M.~Szalay, B.~Sonkoly, G.~Pongracz, and L.~Toka, ``{Sharpening
  Kubernetes for the Edge},'' in \emph{ACM SIGCOMM Posters and Demos}, 2019.

\bibitem{Rossi2020}
F.~Rossi, V.~Cardellini, F.~{Lo Presti}, and M.~Nardelli, ``{Geo-distributed
  efficient deployment of containers with Kubernetes},'' \emph{Comput.
  Commun.}, vol. 159, pp. 161--174, Jun. 2020.

\end{thebibliography}

\vspace{11pt}
\vspace{-3pt}
\begin{IEEEbiography}[{\includegraphics[width=1in,height=1.25in,clip,keepaspectratio]{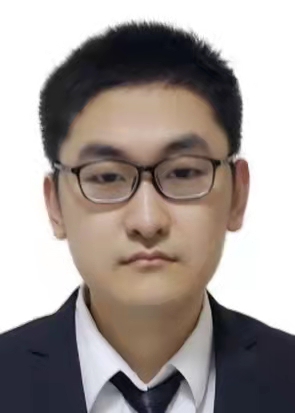}}]{Shihao Shen}
	received his B.S. degree from Tianjin University, China, in 2019. He is currently pursuing the Ph.D. degree with Tianjin University. His research interests include edge computing, cluster scheduling, reinforcement learning, and stochastic optimization.
\end{IEEEbiography}

\vspace{-3pt}
\begin{IEEEbiography}[{\includegraphics[width=1in,height=1.25in,clip,keepaspectratio]{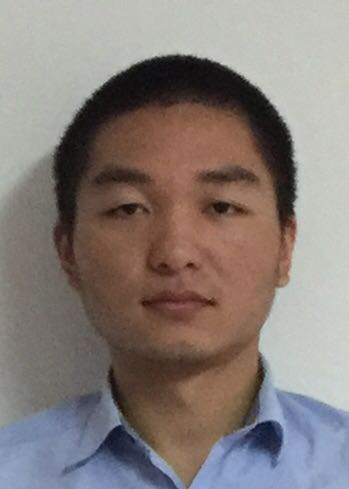}}]{Yiwen Han}
	received the B.S. deqree (with Outstanding Graduates) in communication enqineering from Nanchang University, China, in 2015, and the M.S. degree and the Ph.D. degree (with Outstanding Graduates) from Tianjin University, China, in 2018 and 2022 respectively. He received the National Scholarship of China in both 2016 and 2021. He is currently working on information technology in public safety, and his research interests include edge computing, reinforcement learning, and deep learning.
\end{IEEEbiography}

\vspace{-3pt}
\begin{IEEEbiography}[{\includegraphics[width=1in,height=1.25in,clip,keepaspectratio]{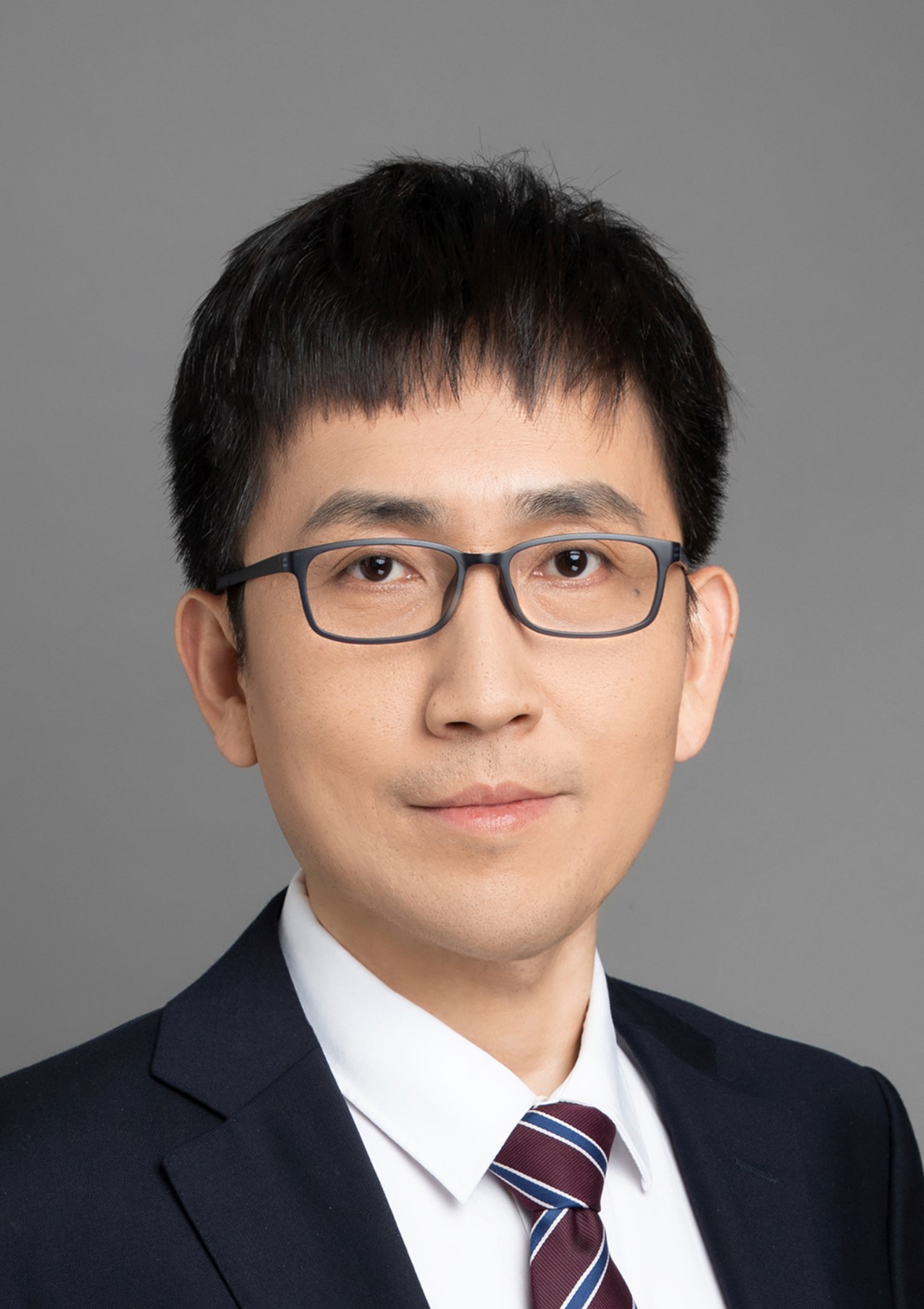}}]{Xiaofei Wang}
	received the B.S. degree from Huazhong University of Science and Technology, China, and received M.S. and Ph.D. degrees from Seoul National University, Seoul, South Korea. He was a Postdoctoral Fellow with The University of British Columbia, Vancouver, Canada, from 2014 to 2016. He is currently a Professor with the College of Intelligence and Computing, Tianjin University, Tianjin, China. Focusing on the research of edge computing, edge intelligence, and edge systems, he has published more than 160 technical papers in IEEE JSAC, TCC, ToN, TWC, IoTJ, COMST, TMM, INFOCOM, ICDCS and so on. He has received the best paper awards of IEEE ICC, ICPADS, and in 2017, he was the recipient of the "IEEE ComSoc Fred W. Ellersick Prize", and in 2022, he received the "IEEE ComSoc Asia-Pacific Outstanding Paper Award".
\end{IEEEbiography}

\vspace{-3pt}
\begin{IEEEbiography}[{\includegraphics[width=1in,height=1.25in,clip,keepaspectratio]{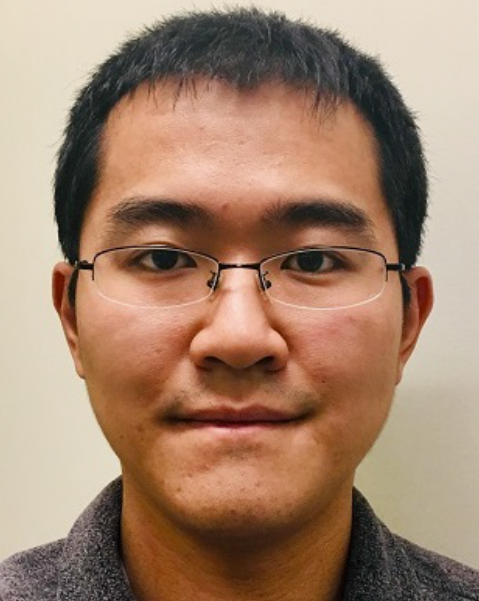}}]{Shiqiang Wang}
	is a Staff Research Scientist at IBM T. J. Watson Research Center, NY, USA. He received his Ph.D. from Imperial College London, United Kingdom, in 2015. His current research focuses on the intersection of distributed computing, machine learning, networking, and optimization, with a broad range of applications including data analytics, edge-based artificial intelligence (Edge AI), Internet of Things (IoT), and future wireless systems. He has made foundational contributions to edge computing and federated learning that generated both academic and industrial impact. Dr. Wang serves as an associate editor of the IEEE Transactions on Mobile Computing and IEEE Transactions on Parallel and Distributed Systems. He received the IEEE Communications Society (ComSoc) Leonard G. Abraham Prize in 2021, IEEE ComSoc Best Young Professional Award in Industry in 2021, IBM Outstanding Technical Achievement Awards (OTAA) in 2019, 2021, and 2022, multiple Invention Achievement Awards from IBM since 2016, Best Paper Finalist of the IEEE International Conference on Image Processing (ICIP) 2019, and Best Student Paper Award of the Network and Information Sciences International Technology Alliance (NIS-ITA) in 2015.
\end{IEEEbiography}
\vspace{-3pt}

\begin{IEEEbiography}[{\includegraphics[width=1in,height=1.25in,clip,keepaspectratio]{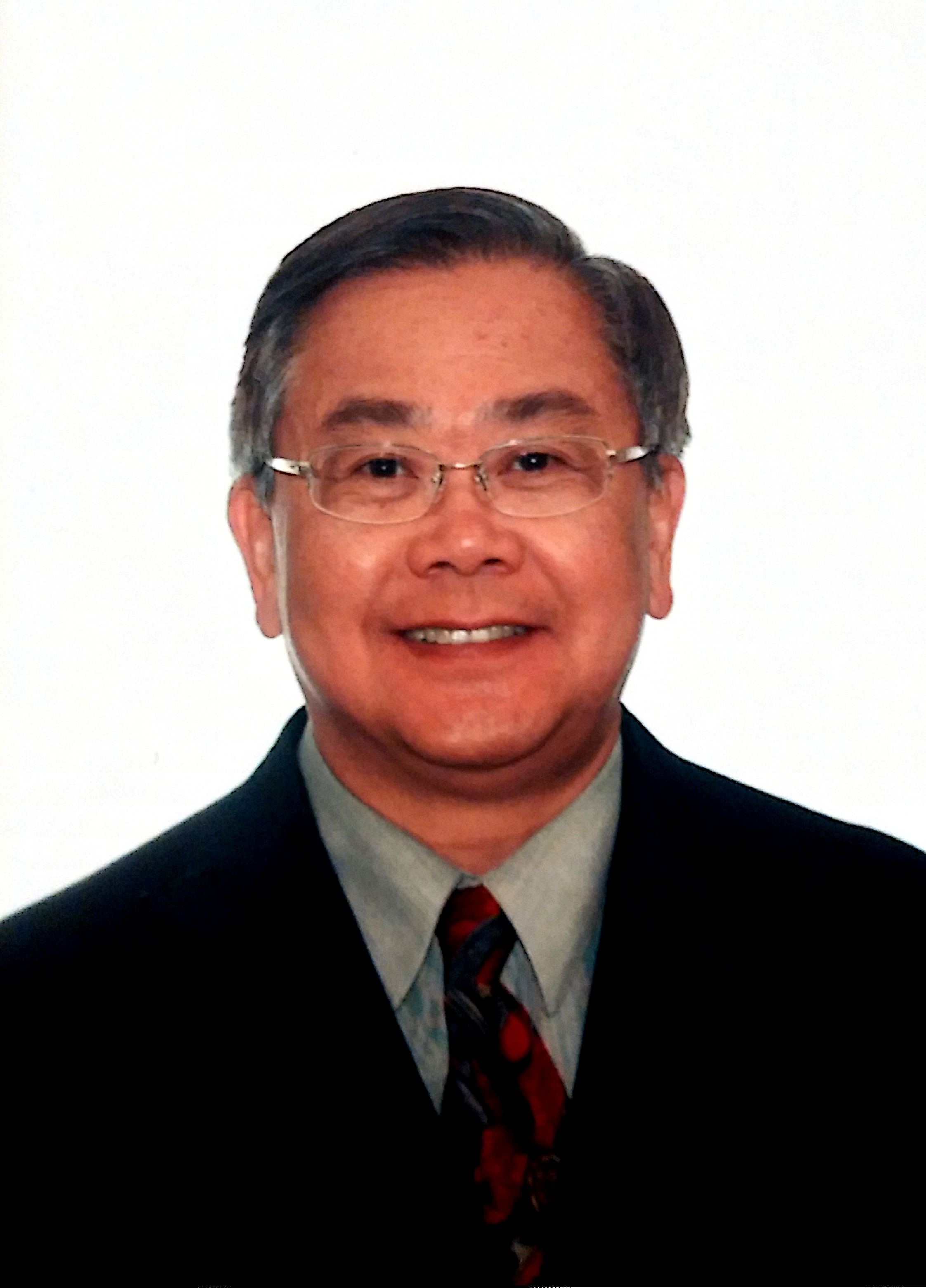}}]{Victor C.M. Leung}
	is a Distinguished Professor of Computer Science and Software Engineering at Shenzhen University, China. He is also an Emeritus Professor of Electrical and Computer Engineering and Director of the Laboratory for Wireless Networks and Mobile Systems at the University of British Columbia (UBC), Canada.  His research is in the broad areas of wireless networks and mobile systems, and he has published widely in these areas. His published works have together attracted more than 50,000 citations. He is named in the current Clarivate Analytics list of “Highly Cited Researchers”. Dr. Leung is serving on the editorial boards of the IEEE Transactions on Green Communications and Networking, IEEE Transactions on Cloud Computing, IEEE Transactions on Computational Social Systems, IEEE Access, and several other journals. He received the 1977 APEBC Gold Medal, 1977-1981 NSERC Postgraduate Scholarships, IEEE Vancouver Section Centennial Award, 2011 UBC Killam Research Prize, 2017 Canadian Award for Telecommunications Research, 2018 IEEE TCGCC Distinguished Technical Achievement Recognition Award, and 2018 ACM MSWiM Reginald Fessenden Award. He co-authored papers that won the 2017 IEEE ComSoc Fred W. Ellersick Prize, 2017 IEEE Systems Journal Best Paper Award, 2018 IEEE CSIM Best Journal Paper Award, and 2019 IEEE TCGCC Best Journal Paper Award.  He is a Life Fellow of IEEE, and a Fellow of the Royal Society of Canada (Academy of Science), Canadian Academy of Engineering, and Engineering Institute of Canada. 
\end{IEEEbiography}

\end{document}